\pdfoutput=1
\documentclass[journal=jacsat,manuscript=article]{achemso}
\SectionNumbersOn

\usepackage[version=3]{mhchem} 
\usepackage{color}
\usepackage{graphics}
\usepackage{natmove}
\usepackage{bm}
\usepackage{float}
\setkeys{acs}{articletitle = true}

\author{Ngoc Thanh Thuy Tran}
\affiliation{Hierachical Green-Energy Materials (Hi-GEM) Research Center, National Cheng Kung University, Tainan 70101, Taiwan}
\author{Duy Khanh Nguyen}
\affiliation{Department of Physics, National Cheng Kung University, Tainan 701, Taiwan}
\author{Shih-Yang Lin}
\affiliation{Department of Physics, National Chung Cheng University, Tainan 701, Taiwan}
\author{Godfrey Gumbs}
\affiliation{Department of Physics and Astronomy, Hunter College of the City University of New York, New York, USA}
\author{Ming Fa-Lin}
\affiliation{Department of Physics, National Cheng Kung University, Tainan 701, Taiwan}
\alsoaffiliation{Quantum Topological Center, National Cheng Kung University, Tainan 701, Taiwan}
\alsoaffiliation{Hierachical Green-Energy Materials (Hi-GEM) Research Center, National Cheng Kung University, Tainan 70101, Taiwan}
\email{mflin@mail.ncku.edu.tw}
\title{Fundamental properties of transition-metals-adsorbed graphene}

\begin{document}
	
	
	\begin{abstract}
	\par\noindent
	The revealing properties of transition metal (T)-doped graphene systems are investigated with the use of the first-principles method. The detailed calculations  cover the bond length, position and height of adatoms, binding energy, atom-dominated band structure, adatom-induced free carrier density as well as energy gap, spin-density distributions, spatial charge distribution, and atom-, orbital- and spin-projected density-of-states (DOS). The magnetic configurations are clearly identified from the total magnetic moments, spin-split energy bands, spin-density distributions and spin-decomposed DOS. Moreover, the single- or multi-orbital hybridizations in T-C, T-T, and C-C bonds can be accurately deduced from the careful analyses of the above-mentioned physical quantities. They are responsible for the optimal geometric structure, the unusual electronic properties, as well as the diverse magnetic properties. All the doped systems are metals except for the low-concentration Ni-doped ones with semiconducting behavior. In contrast, ferromagnetism is exhibited in various Fe/Co-concentrations but only under high Ni-concentrations. Our theoretical predictions are compared with available experimental data, and potential applications are also discussed.			\medskip
		\par
		\noindent \textit{Keywords}: Transition metal, graphene, first-principles, electronic, magnetic.
	\end{abstract}


	\newpage
	\par\noindent
	\section{Introduction} 
	\label{sec1}
	
	\medskip
	\par
	
	More than a decade has elapsed since 2004 when the discovery of few-layer graphene systems was reported. Ever since then, as a consequence of its remarkable properties, graphene has been considered as one of the promising materials that could greatly improve the performance of many products, e.g., capacitors \cite{balci2018electrically,tamura2018capacity}, Li-ion batteries \cite{el2016graphene,raccichini2017critical}, gas sensors \cite{yuan2013graphene,wu2018optical,leenaerts2008adsorption,varghese2015recent}, and spintronic devices \cite{han2014graphene,benitez2018strongly,sierra2018thermoelectric}. Finding ways to greatly diversify the fundamental properties of layered graphene is one of the main-stream topics in physical, chemical and material sciences so  as to expand their potential applications. Available methods include modulation of the number of layers, stacking configuration, guest-atom adsorption/intercalation/substitution, and defects. Significant efforts for creating dramatic transformations of graphene's  fundamental properties have been extensively reported in the literature. So far, chemical modification is considered to be the most efficient technique. The adatom-doped graphene structures have been attracting a considerable amount of theoretical \cite{tran2016chemical,tran2016pi,nakada2011dft,brar2011gate,tran2017coverage}  and experimental \cite{katoch2018transport,gao2010first,dai2010adsorption} attention. The dopant-created chemical bonding will continue to play a critical role in determining the most stable configuration and thus other essential properties. More recently, the transition metal (T)-doped graphene structures have gained a great deal of attention \cite{woo2017temperature,liu2012growth,zanella2008electronic,liu2011bonding}. These systems have been reported to exhibit magnetism and other interesting properties that are promising for spintronic devices and batteries \cite{he2014atomic,cao2010transition,naji2014adsorption}. The aim of our present work is to investigate the transition metal-enriched fundamental properties of monolayer graphene.

	\medskip
	\par

	It is worthy noting that the chemisorption of transition metal adatoms to carbon honeycomb lattice creates a fully modified system without damaging the distinctive one-atom-thick construction itself. Fe- \cite{robertson2013dynamics,he2014atomic,liu2015growth}, Co- \cite{wang2011doping,donati2014tailoring,liu2015growth}, and Ni- \cite{naji2014adsorption,xu2016dft,liu2015growth} doped graphene materials have been successfully produced. In general, there exist two available methods for depositing transition metals onto graphene-based materials.  These are the direct adsorption on graphene sheets \cite{chan2008first,sevinccli2008electronic,virgus2014stability,rigo2009electronic,chan2011gated,power2011magnetization,pi2009electronic} and substitution via vacancy defects  \cite{krasheninnikov2009embedding,santos2010magnetism,lisenkov2012magnetic,boukhvalov2009destruction}. The most stable adsorption sites of Fe/Co/Ni on graphene have been verified using scanning tunneling microscopy (STM) \cite{gyamfi2012orbital,eelbo2013influence,donati2013magnetic}. Additionally, the magnetic moment of Co-absorbed graphene is delicately measured by STM spin-excitation spectroscopy \cite{donati2013magnetic}. From a theoretical point of view, the previous studies are conducted on many adatom types of transition-metal-doped graphene systems \cite{cao2010transition,naji2014adsorption,liu2011bonding},  however, the close relations among the geometrical structure, electronic properties, magnetic configuration, and various adatom concentrations as well as their distributions have not been thoroughly analyzed from the numerical calculations. The multi-orbital hybridization between adatom and graphene, an important issue in understanding the electronic properties modifications, is only slightly mentioned in several studies. Most of them mostly focus on low adatom-concentrations  \cite{liu2011bonding,valencia2010trends} and on investigation of the high-concentration (case 50\% single-side doping) \cite{zanella2008electronic} in the absence of any detailed discussions. That is, the critical mechanisms and pictures for the distinct physical properties are so far lacking.  
		
	\medskip
	\par
	
	In this paper, a theoretical framework is further developed to systematically investigate the orbital hybridizations and spin distributions, which play critical roles in creating the diverse electronic properties and magnetic configurations of the structures under investigation. It should be noticed that the single- or multi-orbital chemical bondings can be revealed from the  atom-dominated energy bands, the spatial charge distributions, and the atom- and orbital-projected DOS. Additionally, the magnetic configurations, namely non-magnetic, ferromagneti, and anti-ferromagnetic, are examined by using spin-split band structures, the magnetic moments, and the spin-decomposed DOS. For example, the electronic properties of oxygen doped monolayer and few-layered graphene structures (finite-gap semiconductors, zero-gap  semiconductors, and semimetals) due to the critical chemical bondings in C-C, C-O, and O-O bonds are indentified by the O-, (C,O)-, and C-dominated bands, the spatial charge density before and after oxygen adsorptions, and the orbital-projected DOS \cite{tran2016pi}. The anti-ferromagnetic configurations of zigzag graphene nanoribbons, being created by the carbon edge's spins, are observed in the spin-degenerate band structure and DOS, and the opposite spin directions across the ribbon centers \cite{nguyen2017fluorination}. This will serve as a first step toward a full understanding of the varied electronic and magnetic properties not only for adatom-doped graphene systems, but also could be further generalized for other one-dimensional (1D) and 2D materials such as silicene-, germanene-, tinene-, MoS$_2$-, and phosphorene-related layered systems. This work  provides accurate results for the diversified properties due to the Fe-, Co-, and Ni-adsorptions with a wide range of adatom concentrations from fully  (100\%; double-side) to low adsorptions (3\%; single-side) through  first-principles calculations by employing the density functional theory (DFT) \cite{gross2013density}. Most important of all,  concise chemical and physical pictures are proposed to account for the presented rich and unique phenomena.
	
	\medskip
	\par
	
	The first-principles calculations on Fe, Co and Ni-adsorbed graphene systems include determining the binding energies, bond lengths, positions and heights of adatoms, atom-dependent band structures, adatom-induced free carrier densities, spatial charge distributions, spin configurations, magnetic moments, and atom-, orbital- and spin-projected DOSs. The meticulous analyses of the calculated results are very useful for fully understanding the cooperative/competitive relations between the honeycomb lattice, the critical orbital hybridizations in carbon-adatom bonds, and the spin configurations. Whether there exist distinct electronic properties (semiconductors or metals) and magnetic configurations (non-magnetism, ferromagnetism, and anti-ferromagnetism) is carefully explored.The above-mentioned results can be verified by experimental measurements, such as, STM \cite{gyamfi2012orbital}, TEM \cite{lee2008growth}, angle resolved photoemission spectroscopy (ARPES) \cite{vilkov2013controlled}, and scanning tunneling spectroscopy \cite{gao2014probing}(STS). Our complete and reliable results should be very helpful in the design and development of potential device applications. 	
	
	\section{Theoretical model}
	\label{sec2}
	
	We note here that our first-principles calculations were performed with the use of the density functional theory through the Vienna {\em ab initio\/} simulation software package (VASP) \cite{kresse1996efficient,kresse1999ultrasoft}.The projector augmented wave method was employed to evaluate the electron-ion interactions \cite{blochl1994projector}, whereas the electron-electron Coulomb interactions belong to the many-particle exchange and correlation energies under the Perdew-Burke-Ernzerhof generalized gradient approximation method \cite{perdew1996generalized}. To carefully explore the chemical adsorption effects on the magnetic properties, the spin configurations are taken into account. The vacuum distance along the z-axis is set to be 15 $\AA$ in order to suppress the van der Waals interactions between two neighboring cells. A plane-wave basis set, with a maximum energy cutoff of 500 eV is available in the calculations of Bloch wave functions. All atomic coordinates are relaxed until the Hellmann-Feynman force on each atom is less than 0.01 eV/$\AA$. The pristine first Brillouin zone is sampled in a Gamma scheme along the two-dimensional periodic direction by $30\times 30\times 1$ k-points for structure relaxations, and then by $100\times 100\times 1$ ${\bf k}$-points for further evaluations on electronic properties. Equivalent ${\bf k}$-point mesh are builded for other enlarged cells depending on their sizes. Furthermore, the van der Waals force, which utilizes the semiemprical DFT-D2 correction of Grimme \cite{grimme2006semiempirical} is very useful in understanding the significant atomic interactions between layers.
	
\medskip
	\par

	A fitting question which might be asked is how many kinds of electronic properties (finite-gap semiconductors, zero-gap semiconductors, semimetals and metals) and magnetic configurations (non-magnetism, ferromagnetism and anti-ferromagnetism) will be verified using the proposed theoretical framework. The answer lies in the way we deal with the delicate atom- and orbital-projected DOS, the atom-dominated energy bands and the spatial charge densities  before and after the adatom adsorptions, which provide much information about the dominating chemical bondings. They could be used to identify the complex multi-orbital hybridizations in C-C, C-T and T-T chemical bonds, being one of the topics we focus on. Such bondings play a critical role in the fundamental properties, accounting for the rich and unique geometric structures and electronic properties of the Fe/Co/Ni-doped graphen systems. Moreover, the spin distributions are important for creating the diverse magnetic configurations. The spin-polarized calculations are utilized to confirm the Fe-, Co- and Ni-enriched magnetic configurations by the use of magnetic moments, spin-split energy bands, spin arrangements, and spin-projected DOS. 
		
	\medskip
	\par
	
	\vspace{5mm}
	\par\noindent
	\section{Geometric, electronic and magnetic properties}
	\label{sec3}
	The various Fe/Co/Ni-adsorption structures, critical multi-orbital hybridizations, significant nonmagnetism/ferromagnetism, and metallic/semiconducting behaviors are worthy of systematic investigation. The typical adatoms, concentrations, and distributions will clearly illustrate diversified phenomena. The up-to-date experimental examinations and potential applications are also discussed in this section.
		
	\medskip
	\par
	
	\subsection{Optimal geometric structures}
	
\begin{figure}[hp]
		\graphicspath{{figure}}
		\centering
		\includegraphics[scale=2.7]{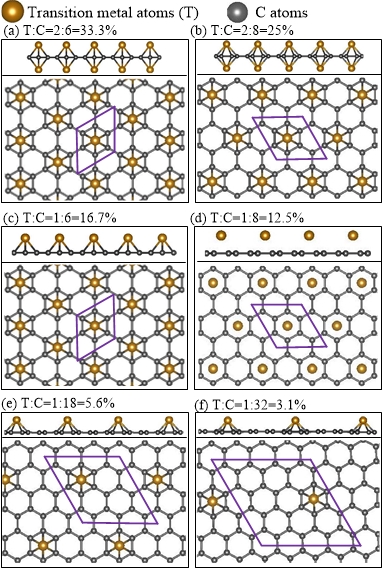}
		\centering\caption{(Color online) Side- and top-view of the geometric structures for Fe/Co/Ni-doped graphene systems under various concentrations and distributions: (a) T:C = 2:6 = 33.3\%${^\ast}$, (b) T:C = 2:8 = 25\%${^\ast}$, (c) T:C = 1:6 = 16.7\%, (d) T:C = 1:8 = 12.5\%, (e) T:C = 1:18 = 5.6\%, and (f) T:C = 1:32 = 3.1\%.}
				\label{FIG:1}
	\end{figure}

	The fundamental properties of transition-metal-adsorbed graphene systems are fully explored under the various adatoms, concentrations and distributions. Based on the completely calculated results, transition metals are preferably adsorbed at the hollow-site compared to the top- and bridge-sites, which is consistent with previous theoretical  predictions \cite{cao2010transition,mao2008density,naji2014adsorption,tang2015adsorption}. The adatom distribution of transition metals on a graphene surface could be examnied by experimental measurement, for which STM is an efficient tool. The hollow-site adsorption of Co and Ni adatoms on graphene has been successfully executed using STM \cite{gyamfi2012orbital,eelbo2013influence,donati2013magnetic}. Specially, it is very hard for STM measurements to verify the single doping site of Fe on graphene at low concentrations due to its small difussion barrier \cite{eelbo2013adatoms}. The main characteristics of optimal geometries, the C-C bond lengths, C-T bond lengths, and heights of the T adatoms relative to the graphene plane, are strongly dependent on the adatom configurations, as obviously indicated in Table 1. Clearly depicted by  the side-view structure in Fig. \ \ref{FIG:1}, Fe/Co/Ni-doped graphene configurations do not generate significant buckling even at high concentrations. That is, the planar honeycomb lattice almost remains planar so that the $\sigma$ bondings due to (2s,2p$_x$,2p$_y$) orbitals of carbon atoms hardly change after the strong chemisorption (discussed later pertaining to Fig. \ \ref{FIG:6}). The nearest C-C bond lengths of passivated C atoms are slightly increased in the specific ranges of 1.43-1.46 $\AA$, reflecting the weak modifications in the $\sigma$ bondings. Regarding the C-T bond lengths and adatom heights (measured from the graphene plane) at low concentrations (Fig. \ \ref{FIG:1}(f)), their magnitudes slightly vary within the range of 2.11-2.13 $\AA$ and 1.51-1.54 $\AA$, respectively. At high concentration, the C-T bond lengths and adatom heights are increased to almost 3.0 $\AA$ at 100\% T-concentrations,  which we discuss later on.
		
	\medskip
	\par

\begin{figure}[hp]
		\graphicspath{{figure}}
		\centering
		\includegraphics[scale=2.3]{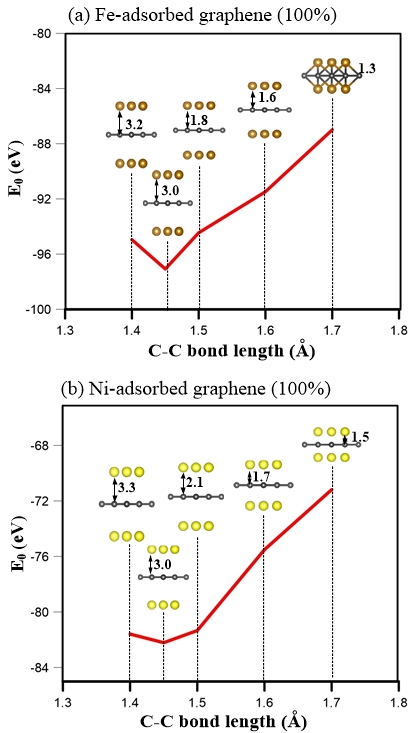}
		\centering\caption{(Color online) Total ground state energy ($E_0$) as a function of nearest-neighbor  C-C bond length for (a) Fe-adsorbed and (b) Ni-adsorbed  graphene for  ful (Fe,Ni)-adsorptions. }
\label{FIG:2}
	\end{figure}

	The binding energy, which characterizes the lowered total ground state energy after the chemical adsorption of transition metal adatoms, is expressed as: $E_b = (E_{sys} - E_{gra} - nE_T)/n$, where $E_{sys}$, $E_{gra}$, and $E_T$ are the total energies, corresponding to the adatom-adsorbed graphene system, the graphene sheet and the isolated transition metal atoms, respectively. Also, $n$ is the number of adatoms per unit cell. In general, based on our calculated results, presented in Table 1, Fe-adsorbed graphene systems are more stable than the Co- and Ni-adsorbed cases at high-concentration, which is in agreement with previous results presented in \cite{liu2014structures,zanella2008electronic,eelbo2013adatoms}. Specifially, for the highest concentrations (100\% for the double-side adsorption), when the C-C bond length is increased from 1.4 $\mbox\AA$ to 1.7$\mbox\AA$ during the optimization procedure, and the height of adatom is significantly to lie in the range of 3.2 $-$ 1.3 $\mbox\AA$ as shown in Fig. \ \ref{FIG:2}. This might be due to the van der Waals interactions between adatom layers and graphene plane \cite{horing2013atom}. Based on the total ground state energy, the larger interlayer distance between an adatom and graphene shows greater stability, in which the corresponding C-C bond length and adatom height of 1.45$\mbox\AA$ and 3.0 $\mbox\AA$ is the most stable. These adatom heights could be verified using transmission electron microscopy (TEM), a useful tool for identifying the interlayer distances of few-layer graphene \cite{lee2008growth}.

	\subsection{Electronic properties and magnetic configurations}
	\medskip
	\par
	
\begin{figure}[hp]
		\graphicspath{{figure}}
		\centering
		\includegraphics[scale=0.45]{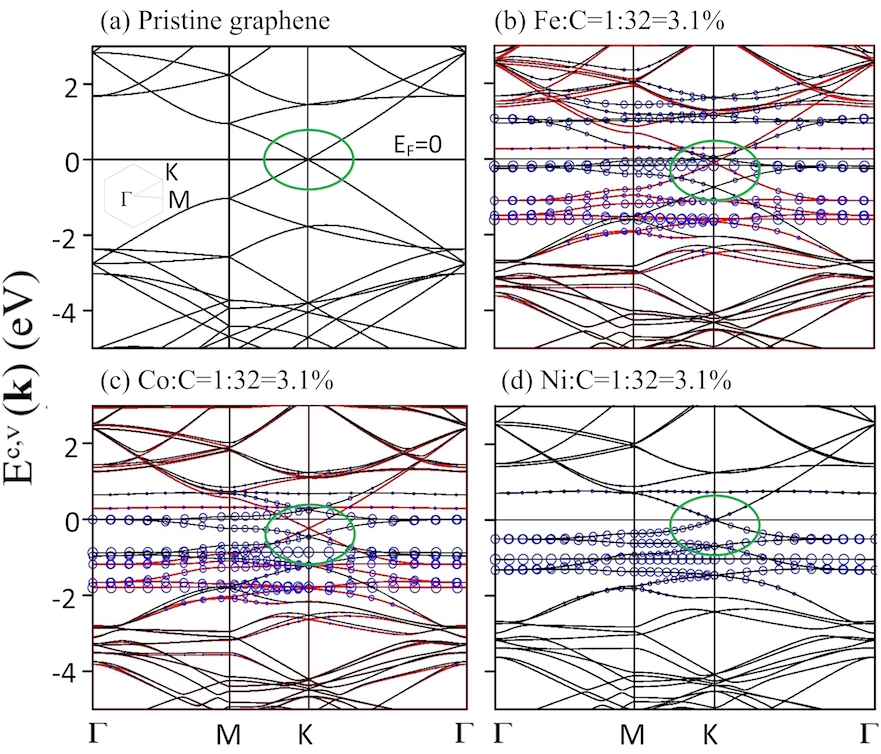}
			\centering\caption{(Color online)  Band structures in a ${4\times\,4}$ supercell (32 C atoms per cell) for (a) monolayer graphene, and (b) Fe-, (c) Co-, and (d) Ni-doped graphene with concentration T:C = 1:32 = 3.1\%. The blue circles represent the contributions from adatoms. The red and black curves correspond to  spin-up and spin-down states, respectively.}
\label{FIG:3}
	\end{figure}

	The 2D band structures along the high-symmetry directions are helpful in examining the main features of the  electronic properties. For pristine monolayer graphene in Fig.\  \ref{FIG:3}(a), there exists a pair of linear valence and conduction bands intersecting at the K/K$^\prime$ point because of the hexagonal symmetry. The intersection is located at the Fermi level ($E_F$) and is usually referred to as a Dirac point. The sp$^2$ orbitals of (2s,2p$_x$,2p$_y$) form very strong covalent $\sigma$-bonds between nearest-neighbor carbon atoms, while the 2p$_z$ orbitals create perpendicular $\pi$ bondings, thereby leading to the low-lying a Dirac cone structure. The $\sigma$ orbitals form the valence bands at the deeper-energy range of ${E^{c,v}<-2.5}$ eV, in which the band edge states initiate from the $\Gamma$ point. On the other hand, the $\pi$ valence and $\pi^\ast$ conduction bands dominate the electronic structures when $|E^{c,v}| <$ 2.5 eV. This is responsible for most of the essential properties. e.g., magnetic quantizations \cite{huang2014feature} and quantum Hall effects \cite{qiao2014quantum}. Regarding the middle-energy electronic states, the saddle points are closely related at the $M$ points (the middle ones between two corners in the hexagonal first Brillouin zone, as shown in the inset of Fig.\  \ref{FIG:3}(a)). Such critical points in the energy-wave-vector space accumulate a lot of states. Therefore, they are expected to exhibit the unusual van Hove singularities for  the DOS (shown in Fig.\  \ref{FIG:7}) and thus the special optical absorption structures \cite{santoso2014tunable,chen2014shift}.

	\medskip
	\par
	
	The low-lying Dirac-cone structure of graphene is dramatically modified after the Fe/Co/Ni-adsorptions. For low-concentration distributions (as presented in Figs. \ref{FIG:3}(b)-\ref{FIG:3}(d)),  Ni-doped graphene (shown in Fig.\ \ref{FIG:3}(d)) differs from the other two adatoms cases, whereby the former is a zero-gap semiconductor and the latter cases are semimetals. This can be observed from the position of the Dirac-cone structure compared with the Fermi level (marked by green ellipses). For Fe and Co-doping configurations  with energy bands in Figs.\ \ref{FIG:3}(b) and \ref{FIG:3}(c), the Fermi level appears at the conduction Dirac cone, indicating the induced free electrons (n-type doping). In case of Ni-doping (\ref{FIG:3}(d)), the Dirac point remains the same at $E_F$ as pristine case. The similarity among Fe, Co, and Ni bands is that their Dirac-cone structures formed by 2p$_z$ orbitals of carbons are isotropic in the range of $|E^{c,v}| <$ 0.5 eV and becomes anisotropic in the presence of band mixing with the adatom-dominated ones. Moreover, there exists spin-up and spin-down energy band splittings near the K and M points except for the case of Ni-doping. There exist the newly created flat s in the energy range of $E \sim -2$ eV to 1 eV, mainly dominated by transition metal and carbon atoms, as shown by the blue circles contribution, in the valence and conduction bands, respectively. This indicates that adatoms make major contributions from low-lying valence electronic states.
		
	\medskip
	\par
	
\begin{figure}[hp]
		\graphicspath{{figure}}
		\centering
		\includegraphics[scale=0.35]{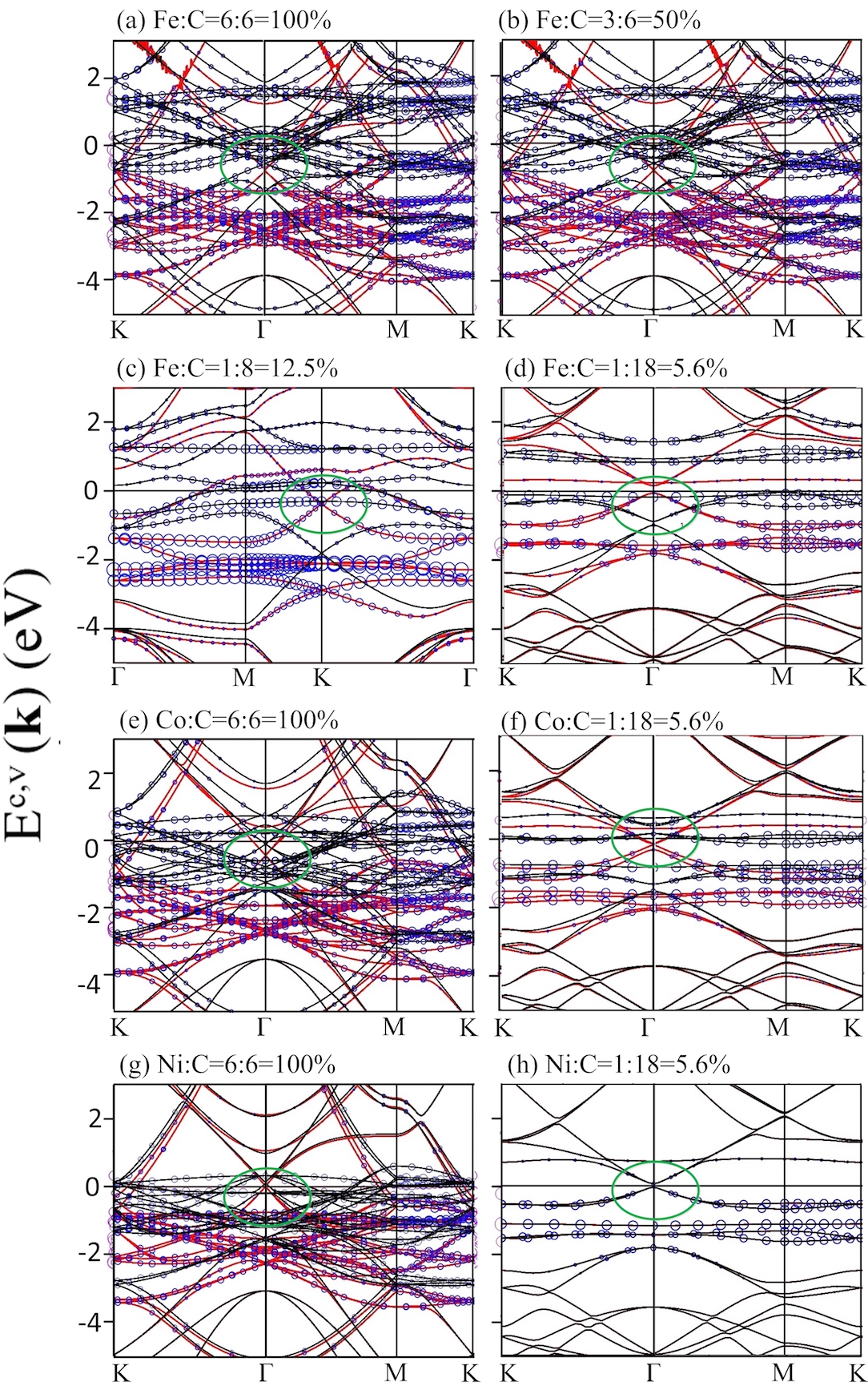}
		\caption{(Color online) Band structures of Fe-, Co- and Ni-doped graphene: (a) Fe:C = 6:6 = 100\% (double-side), (b) Fe:C = 3:6 = 50\% (single-side), (c) Fe:C = 1:8 = 12.5\% (single-side), (d) Fe:C = 1:18 = 5.6\% (single-side), (e) Co:C = 6:6 = 100\% (double-side), (f) Co:C = 1:18 = 5.6\% (single-side), (g) Ni:C = 6:6 = 100\% (double-side), and (h) Ni:C = 1:18 = 5.6\% concentrations. The blue circles correspond to the contributions of adatoms.}
		\label{FIG:4}
	\end{figure}

\begin{figure}[hp]
		\graphicspath{{figure}}
		\centering
		\includegraphics[scale=0.23]{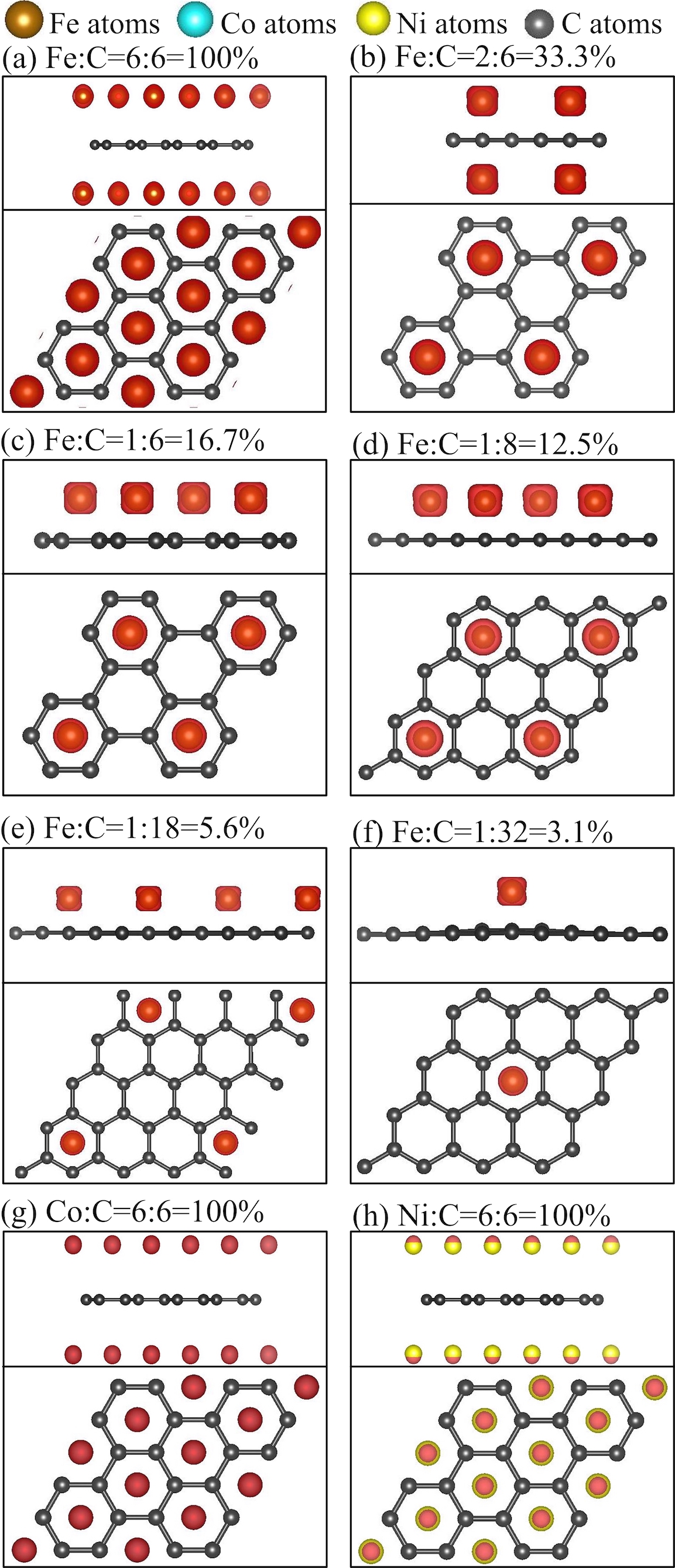}
		\caption{(Color online) The spin-density distributions with top and side views under different concentrations and distributions: (a) Fe:C = 6:6 = 100\%, (b) Fe:C = 2:6 = 33.3\%, (c) Fe:C = 1:6 = 16.7\%, (d) Fe:C = 1:8 = 12.5\%, (e) Fe:C = 1:18 = 5.6\%, (f) Fe:C = 1:32 = 3.1\%, (g) Co:C = 6:6 = 100\%, and (h) Ni:C = 6:6 = 100\%. The red isosurfaces represent the charge density of spin-up configuration.}
		\label{FIG:5}
	\end{figure}

	The significant Fe/Co/Ni-adsorption modifications of the electronic structure can be achieved by tuning the concentration and distribution of adatoms, as clearly presented in Figs.\  \ref{FIG:4}(a)-\ref{FIG:4}(h). There still exists a Dirac-cone structure for even fully absorded configurations, indicating that the $\pi$ bondings are not terminated due to the substantial adatom heights. It is well known that the Dirac cone of graphene is initiated from the K valley. However, it is worth noting that in some unit cells, e.g., corresponding to Figs.\ \ref{FIG:1}(a), \ref{FIG:1}(c) and \ref{FIG:1}(e), it will be transferred to the $\Gamma$ point as a consequence of the of  zone-folding effect. The Dirac point, originally locates at the Fermi level, has a blue shift of $\sim$ 0$-$0.8 eV corressonding to the increasing of T-concentrations. In addition to the anisotropy for $|E| > $0.5 eV around the Dirac point, the Dirac cone structure  quicly becomes parabolic then partially flat bands. Specifically for the case of Fe doping at 5.6\%. This illustrates the semiconducting case arising from the slightly separated Dirac cones. This unusual result id likely due to the partial frustration of the $\pi$ bonding extended in the honeycomb lattice, and we refer to Fig.\ \ref{FIG:4}(d). There exist important differences between the high and low adatom-concentrations. At high T-concentration (larger than 50\%), the electronic dispersions are fully or partially dominated by the adatoms, indicating the T-T and T-C bonding, in a wide energy range  -4 eV $< E^{c,v} <$ 2 eV. Under lower concentrations, the T-dominated bands become narrower. Based on a literature review, the Fe/Co-doped systems are classified as magnetic materials whereas the Ni-doped case is non-magnetic \cite{cao2010transition,johll2014influence,naji2014adsorption}. After detailed and careful calculations, our results show that this is true only for low Ni-concentration.  At high Ni-concentration, e.g. 100\% as shown in Fig.\ \ref{FIG:4}(g), the system also has spin-up and spin-down energy splitting, indicating a ferromagnetic configuration. At lower concentrations, they possess zero magnetic moments, as shown in Table 1, resulting in the absence of spin split bands, as we can see in e.g, Fig.\ \ref{FIG:4}(h). More details related to the magnetic configurations will be discussed later in the spin density subsection (Figs.\ \ref{FIG:5}). On the other hand, the deeper strong $\sigma$ band is hardly affected by the hollow-site chemisorption, only slightly modified at high T-concentrations (partially T-C co-dominated bands). Obviously, there exist a red shift of the $\sigma$ band $\sim$ 0.1$-$0.8 eV away from $E_F$, depending on adatom types and concentrations.
	
	\medskip
	\par
	
 	From an  experimental point of view, the rich features of the band structure could be investigated with the use of ARPES. It is worth noting that ARPES has been employed as a powerful experimental tool to probe the electronic band structure of graphene-related systems, which is not possible in a transport measurement. It has been used to confirm the Dirac cone linear energy dispersion for graphene layers on SiC \cite{sprinkle2009first}. Moreover, ARPES can be used to verify the effects due to  doping. High-resolution ARPES observations on chemically doped graphene of Li \cite{virojanadara2010epitaxial,sugawara2011fabrication} and K  \cite{bostwick2007quasiparticle,ohta2006controlling} atoms have confirmed high density of free electrons in the linear conduction band as evidenced by a red shift of 1.0 $-$1.5 eV of the $\pi$ and $\pi^\ast$ bands. Similarly, the Dirac-cone shift of about 2.6 eV below the Fermi level is observed for graphene-capped Ni silicides \cite{vilkov2013controlled}. The feature-rich bands of transition metal doped graphene, including the red shift of the Dirac point, the degeneracy and anisotropy with different slopes of the Dirac-cone structure, and the adatom-dependent energy bands, can furthermore be examined with ARPES.

\medskip
\par

	The configurations for the spin density distributions could provide additional information regarding their magnetic properties. All the Fe-doped graphene structures display ferromagnetic behavior, as it is clearly illustrated in Figs.\  \ref{FIG:5}(a) and \ref{FIG:5}(h). The spin-up magnetic moments almost dominate the spin-density arrangement, being closely related to the adatom-dominated spin-split energy bands near the Fermi level (Figs.\ \ref{FIG:3}(b) and \ref{FIG:4}(a)-\ref{FIG:4}(d)) and the low-energy DOS of adatom orbitals (Figs. \ref{FIG:7}(a)-\ref{FIG:7}(d)). This is in good agreement with the spatial spin densities accumulated around the adatoms (Figs.\ \ref{FIG:5}(a)-\ref{FIG:5}(h)). Typical magnetic moments per Fe adatoms are about 2-3$\mu_B$ (Bohr magneton). They are sensitive to adatom concentrations and distribution configurations (Table 1). In addition, similar ferromagnetic configurations with magnetic moment $\sim$ 1-2$\mu_B$ per adatom are observed in Co-adsorption graphene (e.g., fully occupied case in Fig.\ \ref{FIG:5}(g)). On the other hand, Ni-doped graphene exhibits an average magnetic moment of 0.83$\mu_B$ per adatom under full and half adsorption cases, whereas the non-magnetic behavior comes to exist at low Ni-concentrations. Obviously, the ferromagnetic moment rapidly decreases in the order of Fe $>$ Co $>$ Ni adatoms. This behavior also occurs for other Fe/Co/Ni related systems, such as Fe/Co/Ni clusters and monolayers on Au/Ir/Pt (111) \cite{bornemann2012trends}, X$-$Pt (X= Fe,Co,Ni) \cite{paudyal2004magnetic}, and Fe/Co/Ni nanowires encapsulated in SiC nanotube \cite{zhang2010comparison}. In addition to the dependence on adatom type, the magnetic moment is almost proportional to the percentage of transition metal adatoms, which might be due to the interaction between adatoms. We note that the transition metal-created magnetic properties on graphene surfaces could be verified using spin-polarized STM \cite{serrate2010imaging,wulfhekel2007spin}.

	\medskip
	\par

\begin{figure}[hp]
		\graphicspath{{figure}}
		\centering
		\includegraphics[scale=1.8]{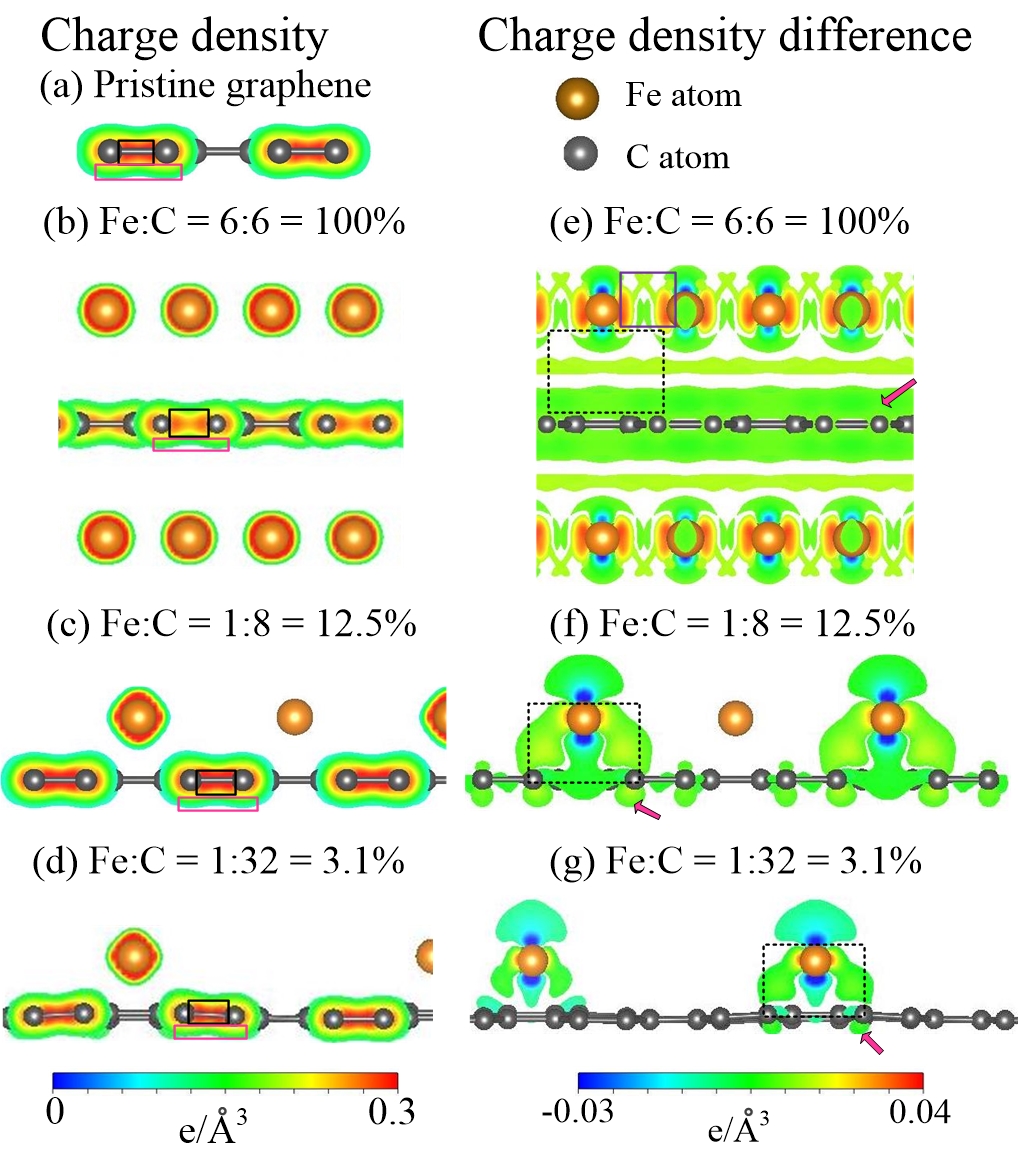}
		\caption{(Color online) The spatial charge densities for  (a) pristine graphene, (b) Fe:C = 6:6 = 100\%, (c) Fe:C = 1:8 = 12.5\%, and (d) Fe:C = 1:32 = 3.1\%. The corresponding charge density differences of Fe-doped graphene are, respectively, shown in (e)-(g).}
\label{FIG:6}
	\end{figure}

\begin{figure}[hp]
		\graphicspath{{figure}}
		\centering
		\includegraphics[scale=0.22]{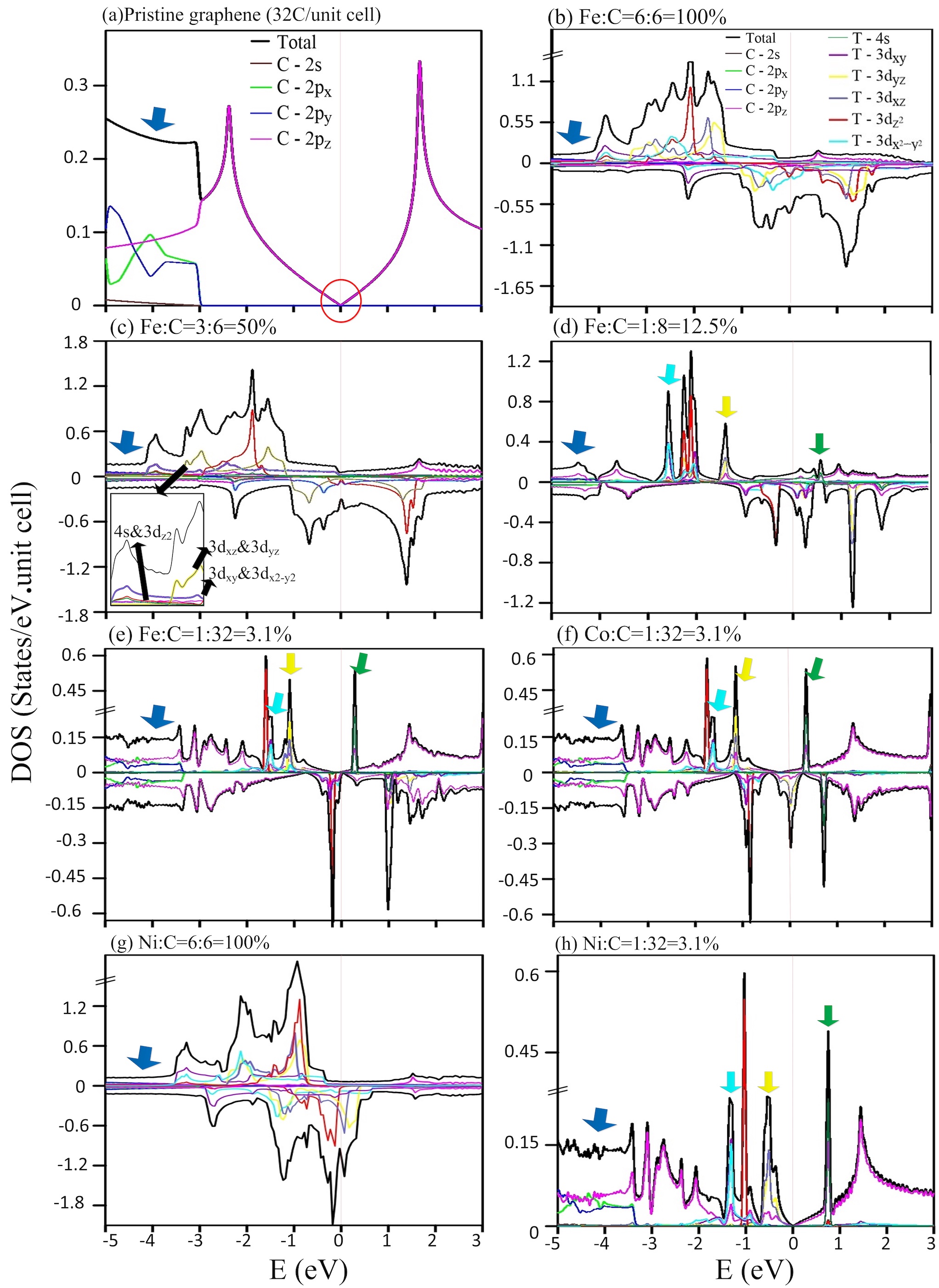}
		\caption{(Color online) Orbital-projected DOS of Fe-, Co- and Ni-doped graphene for the following concentrations and distributions: (a) pristine graphene, (b) Fe:C = 6:6 = 100\%, (c) Fe:C= 3:6 = 50\%, (d) Fe:C = 1:8 = 12.5\%, (e) Fe:C = 1:32 = 3.1\%, (f) Co:C = 1:32 = 3.1\%, (g) Ni:C = 6:6 = 100\%, and (h) Ni:C = 1:32 = 3.1\%. The dim vertical red line indicates the location of the Fermi level.}
\label{FIG:7}
	\end{figure}

 	The charge density $\rho$,  presented in Figs.\ \ref{FIG:6}(a)-\ref{FIG:6}(d), and the charge density difference $\Delta\rho$  in Figs.\ \ref{FIG:6}(e)-\ref{FIG:6}(g)) can provide very useful information regarding the orbital hybridization as well as an understanding of the dramatic changes in the electronic properties. The latter is created by subtracting the charge density of graphene and transition metal atoms from that of the whole system. Apparently, the $\pi$ bonding, which is formed perpendicular and continuously extended on a graphene plane, could survive in all Fe/Co/Ni-adsorbed graphene-related systems, as clearly illustrated by the Dirac cone structure in low-lying energy bands, and for this we refer to Figs.\ \ref{FIG:3} and \ref{FIG:4}. However, for all concentration ranges (from 100\% to 3.1\%), it is still somehow affected by the extra transition metal atoms as evidenced by the pink rectangles in Figs. \ref{FIG:6}(b)-\ref{FIG:6}(d) and the pink arrows in Figs.\ \ref{FIG:6}(e)-\ref{FIG:6}(g). This implies that the T-C bond between (4s,3d) orbitals of the adatom and the 2p$_z$ orbital of carbon. On the other hand, the strong $\sigma$ bondings, with very high charge density between two carbon atoms as seen in the red regions inside the black rectangles of Figs.\ \ref{FIG:6}(b) and \ref{FIG:6}(c) hardly depend on the Fe/Co/Ni adsorptions, corresponding to the slightly modification of $\Delta\rho$ at high adatom concentration (Figs.\  \ref{FIG:6}(e) and \ref{FIG:6}(f)) then vanishing at lower concentrations (e.g, 3.1\% in Figs.\  \ref{FIG:6}(g)). This is responsible for the red shift of the carbon-$\sigma$ bands at deeper energies of co-dominated  (Fe/Co/Ni)-C bands in the middle energy.  Specifically, the observable charge variation within the x-y plane is seen between two neighboring transition metal atoms at sufficiently high concentrations (100\% in Fig.\ \ref{FIG:6}(e)), clearly indicating the multi-orbital hybridization of T-T bonds. In short, there exist T-T, T-C and C-C bonds, for which the former is observed under higher concentrations. 
		
	\medskip
	\par

	The main characteristics of the band structure are directly reflected in the atomic, orbital and spin-projected DOS, which is responsible for the orbital contributions and hybridization in chemical bonds. The form, number and energy of the special structures in the DOS, originating from the band-edge states (the critical points in energy-wave vector space), are very sensitive to chemical bonding and Fe/Co/Ni concentrations.  A  dip, marked by red circles in Fig.\  \ref{FIG:7}(a), prominent symmetric peaks, and shoulder structures come from the almost linear bands, the saddle point and the local extrema points, respectively. For pristine graphene, the $\pi$ and $\pi^\ast$ prominent symmetric peaks due to 2p$_z$-2p$_z$ bondings between C atoms dominate the DOS within the range of $|E^{c,v}| \leq $2.5 eV \cite{neto2009electronic}. Furthermore, the isotropic Dirac cone creates vanishing DOS at the Fermi level. Therefore, monolayer graphene is a zero gap semiconductor (Fig.\  \ref{FIG:7}(b)). In general, these two prominent peaks emerge into several more because of zone-folding effects. The 2p$_z$-dependent DOS of carbon atoms are drastically changed after adsorption. The orbital- and spin-projected DOSs show very complex van Hove singularities and thus identify the significant multi-orbital hybridizations in T-C, C-C and T-T bonds. All the Fe/Co/Ni-adsorbed graphene systems (Figs.\ \ref{FIG:7}(b)-\ref{FIG:7}(g)) present a structure with a dip  due to the Dirac cone at low-lying energies with an obvious blue shift from the Fermi momentum states, except the case of low-concentration Ni  (Fig. \ref{FIG:7}(h))  with vanishing DOS at $E_F$.

	\medskip
	\par

	There exist a considerable number of unusual  T-dominated van Hove singularities, mainly arising from the (4s,3d$_{xy}$,3d$_{yz}$,3d$_{xz}$,3d$_{z^2}$,3d$_{x^2-y^2}$) orbitals of (Fe/Co/Ni) atoms as a result of the T-T bonds in a wide range of $E \sim -2.5$ eV to $E_F$. At higher concentrations, the Fe/Co/Ni-dominated DOS is created with broadened energy range, indicating the T-T bonds of (3d$_{xy}$,3d$_{yz}$,3d$_{xz}$,3d$_{z^2}$,3d$_{x^2-y^2}$) orbitals. Particularly, there are two pairs of orbital hybridizations, (3d$_{yz}$, 3d$_{xz}$) and (3d$_{xy}$, 3d$_{x^2-y^2}$) and multi-hybridizations between them with 3d$_{z^2}$ orbital, in which 3d$_{z^2}$ orbitals contribute to the presence of the prominent peaks. Except for low concentration Ni-doped system (Fig.\  \ref{FIG:7}(h)), the DOS with regard to the conduction $\pi$-states is finite at $E_F$, indicating high induced free carrier density. At low T-concentrations (Figs.\ \ref{FIG:7}(e), \ref{FIG:7}(f), and \ref{FIG:7}(h)), the distorted $\pi$ and $\pi^\ast$ peaks are gradually recovered with spin-up and spin-down splitting for the cases of Fe/Co-adsorbed graphene and without splitting for the ones of Ni adsorption. Part of the $\pi$ peaks of C atoms (pink curve) are co-dominated by the Fe/Co/Ni-created peaks at -3.0 eV $\leq$ E $\leq$ 2 eV, clearly illustrating significant hybridization of (4s,3d$_{xy}$,3d$_{yz}$,3d$_{xz}$,3d$_{z^2}$,3d$_{x^2-y^2}$) orbitals of (Fe/Co/Ni) atoms and 2p$_z$ orbitals of C atoms. Apparently, the co-dominated peaks between the 2p$_z$ orbital of carbon and (3d$_{yz}$, 3d$_{xz}$) and (3d$_{xy}$, 3d$_{x^2-y^2}$) orbitals of transition metals are illustrated by the yellow and cyan color arrows, respectively. Moreover, at very low concentration, such as  5.6\% and 3.1\%, the 2p$_z-$4s hybridization in the C-T bonds (marked by green arrows) appear at the range of $\sim$ 0.2 $-$ 1.2 eV. In addition, the orbital-projected DOS also agrees well with the predictions from band structure and spatial charge densities that the $\sigma$ bonds between carbon atoms almost do not take part in the T-C bond (except at very high concentrations) and only show a slightly red shift from $E_F$ (marked by the deep blue arrows).

	\medskip
	\par

	So far, STS measurement has been considered to be a powerful tool for examining special features in the DOS. The differential conductance (dI/dV) is approximately proportional to the DOS and directly reflects the shape, number and intensity of special structures in the DOS. Additionally,  STS \cite{gao2014probing} and spin-polarized STS could be used to identify the unusual electronic and magnetic configurations, and has served to verify the finite DOS near $E_F$ of graphene flakes on graphite \cite{li2009scanning}, the Fermi level red shift of Bi-doped graphene \cite{chen2015long}. The aforementioned characteristics could also be verified by STS measurements on special features of the DOS, including the dip structure below/near $E_F$, the transition metal-induced extended peaks, and the preserved $\sigma$ shoulder. Measurements with STS on the low and middle-energy peaks are expected to identify the configurations and concentrations of different transition-metals types adsorbed on graphene.	
	
	\medskip
	\par

	Hitherto, graphene-based devices are fast developed, owing to their vital potential. Fe/Co/Ni-doped graphene materials are well-known in the application of gas sensors to absorb harmful gas molecules such as CO, CO$_2$, SO$_2$ and H$_2$S \cite{cortes2018fe,cortes2017adsorption,wanno2014dft,zhang2014dft,cortes2016expanding,tang2015adsorption}. It also could be high performance electrocatalysts for oxygen reduction \cite{chen2015aminothiazole}. Moreover, the performance of Li-ion batteries (LIBs) can be improved by inserting graphene into the anode to achieve morphological optimization and performance \cite{wu2010graphene,hassoun2014advanced,atabaki2013graphene}. Graphene can be added in the manufacture of batteries that are light, durable and suitable for high capacity energy storage, with greatly reduced charging time. It is expected that future studies on adatoms adsorbed graphene will achieve many interesting applications. As reported, transition metal (Fe/C/Ni) oxides have been grown on graphene at hollow sites based on the Kirkendall effect \cite{yu2016growth}. Compared to most transition metals to date, this material shows excellent performance as an anode for LIB with strong reversible capacities and stability at a high current density. When transition metal atoms are adsorbed on graphene, the $\pi$ bonding in pristine graphene is modified due to complex hybridization of 2p$_z$ orbitals of C atoms and the (4s,3d$_{xy}$,3d$_{yz}$,3d$_{xz}$,3d$_{z^2}$,3d$_{x^2-y^2}$) orbitals of (Fe/Co/Ni) adatoms. This leads to the transformation from zero-gap semiconductor (pristine graphene) to  semimetallic systems (doped-systems). The energy difference between the Fermi level and the Dirac point increases with higher adatom concentrations, indicating a larger number of free electrons in the conduction cone. This implies that the electronic conductivity of Fe/Co/Ni-doped graphene is somehow enhanced since the conductivity is proportional to the carrier concentration. Further systematic studies are needed to find out the most suitable absorbers on graphene in order to replace graphite in the anode material of LIBs.

	\section{Concluding remarks}
	\label{sec4}  
	
	We have demonstrated that the chemical and physical properties of transition metal-doped graphene systems are enriched by their configurations and the concentration of adatoms. Their geometrical, electronic and magnetic properties are closely related to the complicated hybridizations between 2p$_z$ orbitals of the C atoms and the (4s,3d$_{xy}$,3d$_{yz}$,3d$_{xz}$,3d$_{z^2}$,3d$_{x^2-y^2}$) orbitals of Fe/Co/Ni. The significant orbital hybridization, the strong $\sigma$  bonding and the weak $\pi$ bonding greatly diversify the band structures. Transition metal adsorbed graphene might exhibit anisotropic and a split Dirac cone structure, as well as (T-T)- and (C,T)-dominated bands at low-lying energies. There exist significant differences between Fe/Co and Ni configurations. Most of our investigated systems display semimetallic behavior except the low-Ni concentration cases. Additionally, the spin distributions are found to be important for creating the diverse magnetic configurations. Specifically, for passivation of the same sublattice, ferromagnetic configurations are revealed for all the Fe/Co-doped systems but only for high Ni-concentration systems. Concerning the DOS, the extra orbital hybridizations are revealed in the DOS as several merged structures when -3 $< E <$2 eV. A decrease in T-concentration results in more low-lying special structures arising from the $\pi$ and $\pi^\ast$ electronic states. 

\medskip
\par
	The theoretical framework proposed in this work will be useful for characterizing other emergent materials, since we have demonstrated its effectiveness for examining the rich and unique properties of low-dimensional structures. Further experimental examiation involving STM, ARPES, and STS, is desirable for the aforementioned main structures of the geometric and electronic properties of Fe/Co/Ni-adsorbed graphene systems. These materials are expected to play a role in potential applications in gas sensors and LIBs'  anode. The present work should serve as a first step towards further investigation into other necessary properties of transition metal-doped graphene for fabrication and potential device applications. In addition, it could pave the way for tight-binding model researchers in determining the important hopping integrals (atomic interactions).

	\newpage
	\par\noindent
	{\bf Acknowledgements}
	
	This work was financially supported by the Hierarchical Green-Energy Materials (Hi- GEM) Research Center, from The Featured Areas Research Center Program within the framework of the Higher Education Sprout Project by the Ministry of Education (MOE) and the Ministry of Science and Technology (MOST 107-3017-F-00-003) in Taiwan. 
	
	\newpage
	\begin{table}[htb]
		\small
		\caption{The calculated nearest passivated C-C and C-T bond lengths, adatom heights,binding energies, energy gaps, and total magnetic moments per adatom of Fe/Co/Ni-absorbed graphene systems. The double-side structures are labeled with *.}
		\label{t1}
		\centering
		\begin{tabular*}{1\textwidth}{@{\extracolsep{\fill}}llllllll} 
			\hline
			Adatom& A:C &\multicolumn{2}{c}{Bond length}& Height &$E_b$ &$E_g$&M$_{tot}$/adatom\\
			&  &C-C (\AA)& C-T (\AA)& (\AA)&(eV)& (eV)&($\mu_B$)\\
			\hline
			Fe& 6:6=100\%$^*$ &1.459 &3.367 &3.034 &-2.56 &0  & 2.9 \\	
			& 3:6=50\% &1.450  &3.297  &2.961  &-2.26  &0  &  2.9 \\	
			& 2:6=33.3\%$^*$ &1.450 &2.215 &1.675 & -1.73 &0  & 2.7\\	
			& 2:8=25\%$^*$ &1.464 &2.238 &1.693 &-1.81 &0  &  2.8 \\
			& 1:6=16.7\% &1.445 &2.211 &1.673 & -2.13&0  &  2.5 \\	
			& 1:8=12.5\% &1.446 &2.195 &1.651 & -1.90 &0  & 2.6 \\	
			& 1:18=5.6\% &1.440 &2.150 &1.589 &-1.26 &0  & 2.0 \\	
			& 1:32=3.1\% &1.438 &2.107 &1.509 &-1.02 &0  & 2.0 \\	
			\hline
			Co& 6:6=100\%$^*$ &1.440 &3.360 &3.036 &-2.31 &0  & 1.9 \\	
			& 3:6=50\% & 1.440 & 3.279 & 2.946 & -1.83 &0  &1.9  \\	
			& 2:6=33.3\%$^*$ &1.450 &2.418 &1.935 & -1.81 &0  & 1.4\\	
			& 2:8=25\%$^*$ &1.467 &2.158 &1.583 &-1.78 &0  &  1.6 \\
			& 1:6=16.7\% &1.444 &2.122 &1.555 & -1.97&0  &  1.3 \\	
			& 1:8=12.5\% &1.440 &2.114 &1.548 &-1.46 &0  & 1.3 \\	
			& 1:18=5.6\% &1.440 &2.121 &1.549 &-1.33 &0  & 1.0\\	
			& 1:32=3.1\% &1.436 &2.119 &1.529 &-1.3 &0  & 1.0 \\	
			\hline	
			Ni& 6:6=100\%$^*$ &1.440 &3.347 &3.017 & -2.26 &0  & 0.8 \\	
			& 3:6=50\% &1.450   & 3.231 & 2.887  &-1.95  &0  &   0.8\\
			& 2:6=33.3\%$^*$ &1.452 &2.189 &1.638 & -1.4 &0  & 0\\	
			& 2:8=25\%$^*$ &1.466 &2.157 &1.582 &-1.29 &0  &  0.1\\
			& 1:6=16.7\% &1.443 &2.129 &1.568 & -1.51&0  &  0 \\	
			& 1:8=12.5\% &1.438 &2.126 &1.566 &-1.41 & 0 & 0\\	
			& 1:18=5.6\% &1.438 &2.130 &1.563 &-1.63 &0  & 0 \\	
			& 1:32=3.1\% &1.436 &2.127 &1.538 &-1.67 &0  & 0 \\	
			\hline
		\end{tabular*}
	\end{table}

	\newpage
	${\centerline {\bf Figure Captions}}$
	
	\begin{itemize}
		\item[Figure 1:] Side- and top-view geometric structures of Fe/Co/Ni-doped graphene for various concentrations and distributions: (a) T:C = 2:6 = 33.3\%, (b) T:C = 2:8 = 25\%, (c) T:C = 1:6 = 16.7\%, (d) T:C = 1:8 = 12.5\%, (e) T:C = 1:18 = 5.6\%, and (f) T:C = 1:32 = 3.1\%.
		\bigskip
		
		\item[Figure 2:] Total ground state energy ($E_0$) and the height between graphene and adatom layer of fully Fe/Ni-adsorptions with respect to various nearest C-C bond lengths.
		\bigskip
		
		\item[Figure 3:] Band structures in a ${4\times\,4}$ supercell (32 C atoms per cell) for (a) monolayer graphene, and the (b) Fe-, (c) Co-, and (d) Ni-doped graphene with concentration T:C = 1:32 = 3.1\%. The blue circles represent the contributions of adatoms. The red and black curves correspond to the spin-up and spin-down states, respectively.
		\bigskip
		
		\item[Figure 4:] Band structures of Fe-, Co- and Ni-doped graphene: (a) Fe:C = 6:6 = 100\% (double-side), (b) Fe:C = 3:6 = 50\% (single-side), (c) Fe:C = 1:8 = 12.5\% (single-side), (d) Fe:C = 1:18 = 5.6\% (single-side), (e) Co:C = 6:6 = 100\% (double-side), (f) Co:C = 1:18 = 5.6\% (single-side), (g) Ni:C = 6:6 = 100\% (double-side), and (h) Ni:C = 1:18 = 5.6\% concentrations. The blue circles correspond to the contributions of adatoms.
		\bigskip
		
		\item[Figure 5:] The spin-density distributions with top and side views under different concentrations and distributions: (a) Fe:C = 6:6 = 100\%, (b) Fe:C = 2:6 = 33.3\%, (c) Fe:C = 1:6 = 16.7\%, (d) Fe:C = 1:8 = 12.5\%, (e) Fe:C = 1:18 = 5.6\%, (f) Fe:C = 1:32 = 3.1\%, (g) Co:C = 6:6 = 100\%, and (h) Ni:C = 6:6 = 100\%. The red isosurfaces represent the charge density of spin-up configuration.
		\bigskip
		
		\item[Figure 6:] The spatial charge densities for: (a) pristine graphene, (b) Fe:C = 6:6 = 100\%, (c) Fe:C = 1:8 = 12.5\%, and (d) Fe:C = 1:32 = 3.1\%. The corresponding charge density differences of Fe-doped graphene are, respectively, shown in (e)-(g).
		
		\item[Figure 7:] Orbital-projected DOS of Fe-, Co- and Ni-doped graphene for concentrations and distributions: (a) pristine graphene, (b) Fe:C = 6:6 = 100\%, (c) Fe:C= 3:6 = 50\%, (d) Fe:C = 1:8 = 12.5\%, (e) Fe:C = 1:32 = 3.1\%, (f) Co:C = 1:32 = 3.1\%, (g) Ni:C = 6:6 = 100\%, and (h) Ni:C = 1:32 = 3.1\%. The dim vertical red line illustrates the Fermi level.
		\bigskip
	\end{itemize}
	
	\newpage
	\bibliography{achemso}

\providecommand{\latin}[1]{#1}
\makeatletter
\providecommand{\doi}
  {\begingroup\let\do\@makeother\dospecials
  \catcode`\{=1 \catcode`\}=2 \doi@aux}
\providecommand{\doi@aux}[1]{\endgroup\texttt{#1}}
\makeatother
\providecommand*\mcitethebibliography{\thebibliography}
\csname @ifundefined\endcsname{endmcitethebibliography}
  {\let\endmcitethebibliography\endthebibliography}{}
\begin{mcitethebibliography}{90}
\providecommand*\natexlab[1]{#1}
\providecommand*\mciteSetBstSublistMode[1]{}
\providecommand*\mciteSetBstMaxWidthForm[2]{}
\providecommand*\mciteBstWouldAddEndPuncttrue
  {\def\EndOfBibitem{\unskip.}}
\providecommand*\mciteBstWouldAddEndPunctfalse
  {\let\EndOfBibitem\relax}
\providecommand*\mciteSetBstMidEndSepPunct[3]{}
\providecommand*\mciteSetBstSublistLabelBeginEnd[3]{}
\providecommand*\EndOfBibitem{}
\mciteSetBstSublistMode{f}
\mciteSetBstMaxWidthForm{subitem}{(\alph{mcitesubitemcount})}
\mciteSetBstSublistLabelBeginEnd
  {\mcitemaxwidthsubitemform\space}
  {\relax}
  {\relax}

\bibitem[Balci \latin{et~al.}(2018)Balci, Kakenov, Karademir, Balci,
  Cakmakyapan, Polat, Caglayan, {\"O}zbay, and Kocabas]{balci2018electrically}
Balci,~O.; Kakenov,~N.; Karademir,~E.; Balci,~S.; Cakmakyapan,~S.;
  Polat,~E.~O.; Caglayan,~H.; {\"O}zbay,~E.; Kocabas,~C. Electrically
  switchable metadevices via graphene. \emph{Sci. Adv.} \textbf{2018},
  \emph{4}, eaao1749\relax
\mciteBstWouldAddEndPuncttrue
\mciteSetBstMidEndSepPunct{\mcitedefaultmidpunct}
{\mcitedefaultendpunct}{\mcitedefaultseppunct}\relax
\EndOfBibitem
\bibitem[Tamura \latin{et~al.}(2018)Tamura, Tomai, Oka, and
  Honma]{tamura2018capacity}
Tamura,~N.; Tomai,~T.; Oka,~N.; Honma,~I. Capacity improvement of the
  carbon-based electrochemical capacitor by zigzag-edge introduced graphene.
  \emph{Appl. Surf. Sci.} \textbf{2018}, \emph{428}, 986--989\relax
\mciteBstWouldAddEndPuncttrue
\mciteSetBstMidEndSepPunct{\mcitedefaultmidpunct}
{\mcitedefaultendpunct}{\mcitedefaultseppunct}\relax
\EndOfBibitem
\bibitem[El-Kady \latin{et~al.}(2016)El-Kady, Shao, and Kaner]{el2016graphene}
El-Kady,~M.~F.; Shao,~Y.; Kaner,~R.~B. Graphene for batteries, supercapacitors
  and beyond. \emph{Nat. Rev. Mater.} \textbf{2016}, \emph{1}, 16033\relax
\mciteBstWouldAddEndPuncttrue
\mciteSetBstMidEndSepPunct{\mcitedefaultmidpunct}
{\mcitedefaultendpunct}{\mcitedefaultseppunct}\relax
\EndOfBibitem
\bibitem[Raccichini \latin{et~al.}(2017)Raccichini, Varzi, Wei, and
  Passerini]{raccichini2017critical}
Raccichini,~R.; Varzi,~A.; Wei,~D.; Passerini,~S. Critical Insight into the
  Relentless Progression Toward Graphene and Graphene-Containing Materials for
  Lithium-Ion Battery Anodes. \emph{Adv. Mater.} \textbf{2017}, \emph{29},
  1603421\relax
\mciteBstWouldAddEndPuncttrue
\mciteSetBstMidEndSepPunct{\mcitedefaultmidpunct}
{\mcitedefaultendpunct}{\mcitedefaultseppunct}\relax
\EndOfBibitem
\bibitem[Yuan and Shi(2013)Yuan, and Shi]{yuan2013graphene}
Yuan,~W.; Shi,~G. Graphene-based gas sensors. \emph{J. Mater. Chem. A}
  \textbf{2013}, \emph{1}, 10078--10091\relax
\mciteBstWouldAddEndPuncttrue
\mciteSetBstMidEndSepPunct{\mcitedefaultmidpunct}
{\mcitedefaultendpunct}{\mcitedefaultseppunct}\relax
\EndOfBibitem
\bibitem[Wu \latin{et~al.}(2018)Wu, Yao, Yu, and Rao]{wu2018optical}
Wu,~Y.; Yao,~B.; Yu,~C.; Rao,~Y. Optical graphene gas sensors based on
  microfibers: A review. \emph{Sensors} \textbf{2018}, \emph{18}, 941\relax
\mciteBstWouldAddEndPuncttrue
\mciteSetBstMidEndSepPunct{\mcitedefaultmidpunct}
{\mcitedefaultendpunct}{\mcitedefaultseppunct}\relax
\EndOfBibitem
\bibitem[Leenaerts \latin{et~al.}(2008)Leenaerts, Partoens, and
  Peeters]{leenaerts2008adsorption}
Leenaerts,~O.; Partoens,~B.; Peeters,~F. Adsorption of H 2 O, N H 3, CO, N O 2,
  and NO on graphene: A first-principles study. \emph{Phys. Rev. B}
  \textbf{2008}, \emph{77}, 125416\relax
\mciteBstWouldAddEndPuncttrue
\mciteSetBstMidEndSepPunct{\mcitedefaultmidpunct}
{\mcitedefaultendpunct}{\mcitedefaultseppunct}\relax
\EndOfBibitem
\bibitem[Varghese \latin{et~al.}(2015)Varghese, Lonkar, Singh, Swaminathan, and
  Abdala]{varghese2015recent}
Varghese,~S.~S.; Lonkar,~S.; Singh,~K.; Swaminathan,~S.; Abdala,~A. Recent
  advances in graphene based gas sensors. \emph{Sens. Actuator B-Chem.}
  \textbf{2015}, \emph{218}, 160--183\relax
\mciteBstWouldAddEndPuncttrue
\mciteSetBstMidEndSepPunct{\mcitedefaultmidpunct}
{\mcitedefaultendpunct}{\mcitedefaultseppunct}\relax
\EndOfBibitem
\bibitem[Han \latin{et~al.}(2014)Han, Kawakami, Gmitra, and
  Fabian]{han2014graphene}
Han,~W.; Kawakami,~R.~K.; Gmitra,~M.; Fabian,~J. Graphene spintronics.
  \emph{Nat. Nanotechnol.} \textbf{2014}, \emph{9}, 794\relax
\mciteBstWouldAddEndPuncttrue
\mciteSetBstMidEndSepPunct{\mcitedefaultmidpunct}
{\mcitedefaultendpunct}{\mcitedefaultseppunct}\relax
\EndOfBibitem
\bibitem[Ben{\'\i}tez \latin{et~al.}(2018)Ben{\'\i}tez, Sierra, Torres,
  Arrighi, Bonell, Costache, and Valenzuela]{benitez2018strongly}
Ben{\'\i}tez,~L.~A.; Sierra,~J.~F.; Torres,~W.~S.; Arrighi,~A.; Bonell,~F.;
  Costache,~M.~V.; Valenzuela,~S.~O. Strongly anisotropic spin relaxation in
  graphene--transition metal dichalcogenide heterostructures at room
  temperature. \emph{Nat. Phys.} \textbf{2018}, \emph{14}, 303\relax
\mciteBstWouldAddEndPuncttrue
\mciteSetBstMidEndSepPunct{\mcitedefaultmidpunct}
{\mcitedefaultendpunct}{\mcitedefaultseppunct}\relax
\EndOfBibitem
\bibitem[Sierra \latin{et~al.}(2018)Sierra, Neumann, Cuppens, Raes, Costache,
  and Valenzuela]{sierra2018thermoelectric}
Sierra,~J.~F.; Neumann,~I.; Cuppens,~J.; Raes,~B.; Costache,~M.~V.;
  Valenzuela,~S.~O. Thermoelectric spin voltage in graphene. \emph{Nat.
  Nanotechnol.} \textbf{2018}, \emph{13}, 107\relax
\mciteBstWouldAddEndPuncttrue
\mciteSetBstMidEndSepPunct{\mcitedefaultmidpunct}
{\mcitedefaultendpunct}{\mcitedefaultseppunct}\relax
\EndOfBibitem
\bibitem[Tran \latin{et~al.}(2016)Tran, Lin, Lin, and Lin]{tran2016chemical}
Tran,~N. T.~T.; Lin,~S.~Y.; Lin,~Y.~T.; Lin,~M.~F. Chemical bonding-induced
  rich electronic properties of oxygen adsorbed few-layer graphenes.
  \emph{Phys. Chem. Chem. Phys.} \textbf{2016}, \emph{18}, 4000--4007\relax
\mciteBstWouldAddEndPuncttrue
\mciteSetBstMidEndSepPunct{\mcitedefaultmidpunct}
{\mcitedefaultendpunct}{\mcitedefaultseppunct}\relax
\EndOfBibitem
\bibitem[Tran \latin{et~al.}(2016)Tran, Lin, Glukhova, and Lin]{tran2016pi}
Tran,~N. T.~T.; Lin,~S.~Y.; Glukhova,~O.~E.; Lin,~M.~F. $\pi$-Bonding-dominated
  energy gaps in graphene oxide. \emph{RSC Adv.} \textbf{2016}, \emph{6},
  24458--24463\relax
\mciteBstWouldAddEndPuncttrue
\mciteSetBstMidEndSepPunct{\mcitedefaultmidpunct}
{\mcitedefaultendpunct}{\mcitedefaultseppunct}\relax
\EndOfBibitem
\bibitem[Nakada and Ishii(2011)Nakada, and Ishii]{nakada2011dft}
Nakada,~K.; Ishii,~A. Migration of adatom adsorption on graphene using DFT
  calculation. \emph{Solid State Commun.} \textbf{2011}, \emph{151},
  13--16\relax
\mciteBstWouldAddEndPuncttrue
\mciteSetBstMidEndSepPunct{\mcitedefaultmidpunct}
{\mcitedefaultendpunct}{\mcitedefaultseppunct}\relax
\EndOfBibitem
\bibitem[Brar \latin{et~al.}(2011)Brar, Decker, Solowan, Wang, Maserati, Chan,
  Lee, Girit, Zettl, Louie, \latin{et~al.} others]{brar2011gate}
Brar,~V.~W.; Decker,~R.; Solowan,~H.-M.; Wang,~Y.; Maserati,~L.; Chan,~K.~T.;
  Lee,~H.; Girit,~{\c{C}}.~O.; Zettl,~A.; Louie,~S.~G., \latin{et~al.}
  Gate-controlled ionization and screening of cobalt adatoms on a graphene
  surface. \emph{Nat. Phys.} \textbf{2011}, \emph{7}, 43--47\relax
\mciteBstWouldAddEndPuncttrue
\mciteSetBstMidEndSepPunct{\mcitedefaultmidpunct}
{\mcitedefaultendpunct}{\mcitedefaultseppunct}\relax
\EndOfBibitem
\bibitem[Tran \latin{et~al.}(2017)Tran, Nguyen, Glukhova, and
  Lin]{tran2017coverage}
Tran,~N. T.~T.; Nguyen,~D.~K.; Glukhova,~O.~E.; Lin,~M.-F. Coverage-dependent
  essential properties of halogenated graphene: A DFT study. \emph{Sci. Rep.}
  \textbf{2017}, \emph{7}, 17858\relax
\mciteBstWouldAddEndPuncttrue
\mciteSetBstMidEndSepPunct{\mcitedefaultmidpunct}
{\mcitedefaultendpunct}{\mcitedefaultseppunct}\relax
\EndOfBibitem
\bibitem[Katoch \latin{et~al.}(2018)Katoch, Zhu, Kochan, Singh, Fabian, and
  Kawakami]{katoch2018transport}
Katoch,~J.; Zhu,~T.; Kochan,~D.; Singh,~S.; Fabian,~J.; Kawakami,~R.~K.
  Transport Spectroscopy of Sublattice-Resolved Resonant Scattering in
  Hydrogen-Doped Bilayer Graphene. \emph{Phys. Rev. Lett.} \textbf{2018},
  \emph{121}, 136801\relax
\mciteBstWouldAddEndPuncttrue
\mciteSetBstMidEndSepPunct{\mcitedefaultmidpunct}
{\mcitedefaultendpunct}{\mcitedefaultseppunct}\relax
\EndOfBibitem
\bibitem[Gao \latin{et~al.}(2010)Gao, Zhou, Lu, Fa, and Chen]{gao2010first}
Gao,~H.; Zhou,~J.; Lu,~M.; Fa,~W.; Chen,~Y. First-principles study of the IVA
  group atoms adsorption on graphene. \emph{J. Appl. Phys.} \textbf{2010},
  \emph{107}, 114311\relax
\mciteBstWouldAddEndPuncttrue
\mciteSetBstMidEndSepPunct{\mcitedefaultmidpunct}
{\mcitedefaultendpunct}{\mcitedefaultseppunct}\relax
\EndOfBibitem
\bibitem[Dai and Yuan(2010)Dai, and Yuan]{dai2010adsorption}
Dai,~J.; Yuan,~J. Adsorption of molecular oxygen on doped graphene: Atomic,
  electronic, and magnetic properties. \emph{Phys. Rev. B} \textbf{2010},
  \emph{81}, 165414\relax
\mciteBstWouldAddEndPuncttrue
\mciteSetBstMidEndSepPunct{\mcitedefaultmidpunct}
{\mcitedefaultendpunct}{\mcitedefaultseppunct}\relax
\EndOfBibitem
\bibitem[Woo \latin{et~al.}(2017)Woo, Hemmatiyan, Morrison, Rathnayaka,
  Lyuksyutov, and Naugle]{woo2017temperature}
Woo,~S.; Hemmatiyan,~S.; Morrison,~T.; Rathnayaka,~K.; Lyuksyutov,~I.;
  Naugle,~D. Temperature-dependent transport properties of graphene decorated
  by alkali metal adatoms (Li, K). \emph{Appl. Phys. Lett.} \textbf{2017},
  \emph{111}, 263502\relax
\mciteBstWouldAddEndPuncttrue
\mciteSetBstMidEndSepPunct{\mcitedefaultmidpunct}
{\mcitedefaultendpunct}{\mcitedefaultseppunct}\relax
\EndOfBibitem
\bibitem[Liu \latin{et~al.}(2012)Liu, Hupalo, Wang, Lu, Thiel, Ho, and
  Tringides]{liu2012growth}
Liu,~X.; Hupalo,~M.; Wang,~C.-Z.; Lu,~W.-C.; Thiel,~P.~A.; Ho,~K.-M.;
  Tringides,~M.~C. Growth morphology and thermal stability of metal islands on
  graphene. \emph{Phys. Rev. B} \textbf{2012}, \emph{86}, 081414\relax
\mciteBstWouldAddEndPuncttrue
\mciteSetBstMidEndSepPunct{\mcitedefaultmidpunct}
{\mcitedefaultendpunct}{\mcitedefaultseppunct}\relax
\EndOfBibitem
\bibitem[Zanella \latin{et~al.}(2008)Zanella, Fagan, Mota, and
  Fazzio]{zanella2008electronic}
Zanella,~I.; Fagan,~S.~B.; Mota,~R.; Fazzio,~A. Electronic and magnetic
  properties of Ti and Fe on graphene. \emph{J. Phys. Chem. C} \textbf{2008},
  \emph{112}, 9163--9167\relax
\mciteBstWouldAddEndPuncttrue
\mciteSetBstMidEndSepPunct{\mcitedefaultmidpunct}
{\mcitedefaultendpunct}{\mcitedefaultseppunct}\relax
\EndOfBibitem
\bibitem[Liu \latin{et~al.}(2011)Liu, Wang, Yao, Lu, Hupalo, Tringides, and
  Ho]{liu2011bonding}
Liu,~X.; Wang,~C.; Yao,~Y.; Lu,~W.; Hupalo,~M.; Tringides,~M.; Ho,~K. Bonding
  and charge transfer by metal adatom adsorption on graphene. \emph{Phys. Rev.
  B} \textbf{2011}, \emph{83}, 235411\relax
\mciteBstWouldAddEndPuncttrue
\mciteSetBstMidEndSepPunct{\mcitedefaultmidpunct}
{\mcitedefaultendpunct}{\mcitedefaultseppunct}\relax
\EndOfBibitem
\bibitem[He \latin{et~al.}(2014)He, He, Robertson, Kirkland, Kim, Ihm, Yoon,
  Lee, and Warner]{he2014atomic}
He,~Z.; He,~K.; Robertson,~A.~W.; Kirkland,~A.~I.; Kim,~D.; Ihm,~J.; Yoon,~E.;
  Lee,~G.-D.; Warner,~J.~H. Atomic structure and dynamics of metal dopant pairs
  in graphene. \emph{Nano Lett.} \textbf{2014}, \emph{14}, 3766--3772\relax
\mciteBstWouldAddEndPuncttrue
\mciteSetBstMidEndSepPunct{\mcitedefaultmidpunct}
{\mcitedefaultendpunct}{\mcitedefaultseppunct}\relax
\EndOfBibitem
\bibitem[Cao \latin{et~al.}(2010)Cao, Wu, Jiang, and Cheng]{cao2010transition}
Cao,~C.; Wu,~M.; Jiang,~J.; Cheng,~H.-P. Transition metal adatom and dimer
  adsorbed on graphene: Induced magnetization and electronic structures.
  \emph{Phys. Rev. B} \textbf{2010}, \emph{81}, 205424\relax
\mciteBstWouldAddEndPuncttrue
\mciteSetBstMidEndSepPunct{\mcitedefaultmidpunct}
{\mcitedefaultendpunct}{\mcitedefaultseppunct}\relax
\EndOfBibitem
\bibitem[Naji \latin{et~al.}(2014)Naji, Belhaj, Labrim, Bhihi, Benyoussef, and
  El~Kenz]{naji2014adsorption}
Naji,~S.; Belhaj,~A.; Labrim,~H.; Bhihi,~M.; Benyoussef,~A.; El~Kenz,~A.
  Adsorption of Co and Ni on graphene with a double hexagonal symmetry:
  electronic and magnetic properties. \emph{J. Phys. Chem. C} \textbf{2014},
  \emph{118}, 4924--4929\relax
\mciteBstWouldAddEndPuncttrue
\mciteSetBstMidEndSepPunct{\mcitedefaultmidpunct}
{\mcitedefaultendpunct}{\mcitedefaultseppunct}\relax
\EndOfBibitem
\bibitem[Robertson \latin{et~al.}(2013)Robertson, Montanari, He, Kim, Allen,
  Wu, Olivier, Neethling, Harrison, Kirkland, \latin{et~al.}
  others]{robertson2013dynamics}
Robertson,~A.~W.; Montanari,~B.; He,~K.; Kim,~J.; Allen,~C.~S.; Wu,~Y.~A.;
  Olivier,~J.; Neethling,~J.; Harrison,~N.; Kirkland,~A.~I., \latin{et~al.}
  Dynamics of single Fe atoms in graphene vacancies. \emph{Nano Lett.}
  \textbf{2013}, \emph{13}, 1468--1475\relax
\mciteBstWouldAddEndPuncttrue
\mciteSetBstMidEndSepPunct{\mcitedefaultmidpunct}
{\mcitedefaultendpunct}{\mcitedefaultseppunct}\relax
\EndOfBibitem
\bibitem[Liu \latin{et~al.}(2015)Liu, Han, Evans, Engstfeld, Behm, Tringides,
  Hupalo, Lin, Huang, Ho, \latin{et~al.} others]{liu2015growth}
Liu,~X.; Han,~Y.; Evans,~J.~W.; Engstfeld,~A.~K.; Behm,~R.~J.;
  Tringides,~M.~C.; Hupalo,~M.; Lin,~H.-Q.; Huang,~L.; Ho,~K.-M.,
  \latin{et~al.}  Growth morphology and properties of metals on graphene.
  \emph{Progress in Surface Science} \textbf{2015}, \emph{90}, 397--443\relax
\mciteBstWouldAddEndPuncttrue
\mciteSetBstMidEndSepPunct{\mcitedefaultmidpunct}
{\mcitedefaultendpunct}{\mcitedefaultseppunct}\relax
\EndOfBibitem
\bibitem[Wang \latin{et~al.}(2011)Wang, \latin{et~al.} others]{wang2011doping}
Wang,~H., \latin{et~al.}  Doping monolayer graphene with single atom
  substitutions. \emph{Nano Lett.} \textbf{2011}, \emph{12}, 141--144\relax
\mciteBstWouldAddEndPuncttrue
\mciteSetBstMidEndSepPunct{\mcitedefaultmidpunct}
{\mcitedefaultendpunct}{\mcitedefaultseppunct}\relax
\EndOfBibitem
\bibitem[Donati \latin{et~al.}(2014)Donati, Gragnaniello, Cavallin, Natterer,
  Dubout, Pivetta, Patthey, Dreiser, Piamonteze, Rusponi, \latin{et~al.}
  others]{donati2014tailoring}
Donati,~F.; Gragnaniello,~L.; Cavallin,~A.; Natterer,~F.; Dubout,~Q.;
  Pivetta,~M.; Patthey,~F.; Dreiser,~J.; Piamonteze,~C.; Rusponi,~S.,
  \latin{et~al.}  Tailoring the magnetism of Co atoms on graphene through
  substrate hybridization. \emph{Phys. Rev. Lett.} \textbf{2014}, \emph{113},
  177201\relax
\mciteBstWouldAddEndPuncttrue
\mciteSetBstMidEndSepPunct{\mcitedefaultmidpunct}
{\mcitedefaultendpunct}{\mcitedefaultseppunct}\relax
\EndOfBibitem
\bibitem[Xu \latin{et~al.}(2016)Xu, Li, Xu, Xu, and Zhao]{xu2016dft}
Xu,~X.-Y.; Li,~J.; Xu,~H.; Xu,~X.; Zhao,~C. DFT investigation of Ni-doped
  graphene: catalytic ability to CO oxidation. \emph{New J. Chem.}
  \textbf{2016}, \emph{40}, 9361--9369\relax
\mciteBstWouldAddEndPuncttrue
\mciteSetBstMidEndSepPunct{\mcitedefaultmidpunct}
{\mcitedefaultendpunct}{\mcitedefaultseppunct}\relax
\EndOfBibitem
\bibitem[Chan \latin{et~al.}(2008)Chan, Neaton, and Cohen]{chan2008first}
Chan,~K.~T.; Neaton,~J.; Cohen,~M.~L. First-principles study of metal adatom
  adsorption on graphene. \emph{Phys. Rev. B} \textbf{2008}, \emph{77},
  235430\relax
\mciteBstWouldAddEndPuncttrue
\mciteSetBstMidEndSepPunct{\mcitedefaultmidpunct}
{\mcitedefaultendpunct}{\mcitedefaultseppunct}\relax
\EndOfBibitem
\bibitem[Sevin{\c{c}}li \latin{et~al.}(2008)Sevin{\c{c}}li, Topsakal, Durgun,
  and Ciraci]{sevinccli2008electronic}
Sevin{\c{c}}li,~H.; Topsakal,~M.; Durgun,~E.; Ciraci,~S. Electronic and
  magnetic properties of 3d transition-metal atom adsorbed graphene and
  graphene nanoribbons. \emph{Phys. Rev. B} \textbf{2008}, \emph{77},
  195434\relax
\mciteBstWouldAddEndPuncttrue
\mciteSetBstMidEndSepPunct{\mcitedefaultmidpunct}
{\mcitedefaultendpunct}{\mcitedefaultseppunct}\relax
\EndOfBibitem
\bibitem[Virgus \latin{et~al.}(2014)Virgus, Purwanto, Krakauer, and
  Zhang]{virgus2014stability}
Virgus,~Y.; Purwanto,~W.; Krakauer,~H.; Zhang,~S. Stability, energetics, and
  magnetic states of cobalt adatoms on graphene. \emph{Phys. Rev. Lett.}
  \textbf{2014}, \emph{113}, 175502\relax
\mciteBstWouldAddEndPuncttrue
\mciteSetBstMidEndSepPunct{\mcitedefaultmidpunct}
{\mcitedefaultendpunct}{\mcitedefaultseppunct}\relax
\EndOfBibitem
\bibitem[Rigo \latin{et~al.}(2009)Rigo, Martins, da~Silva, Fazzio, and
  Miwa]{rigo2009electronic}
Rigo,~V.; Martins,~T.~B.; da~Silva,~A.~J.; Fazzio,~A.; Miwa,~R.~H. Electronic,
  structural, and transport properties of Ni-doped graphene nanoribbons.
  \emph{Phys. Rev. B} \textbf{2009}, \emph{79}, 075435\relax
\mciteBstWouldAddEndPuncttrue
\mciteSetBstMidEndSepPunct{\mcitedefaultmidpunct}
{\mcitedefaultendpunct}{\mcitedefaultseppunct}\relax
\EndOfBibitem
\bibitem[Chan \latin{et~al.}(2011)Chan, Lee, and Cohen]{chan2011gated}
Chan,~K.~T.; Lee,~H.; Cohen,~M.~L. Gated adatoms on graphene studied with
  first-principles calculations. \emph{Phys. Rev. B} \textbf{2011}, \emph{83},
  035405\relax
\mciteBstWouldAddEndPuncttrue
\mciteSetBstMidEndSepPunct{\mcitedefaultmidpunct}
{\mcitedefaultendpunct}{\mcitedefaultseppunct}\relax
\EndOfBibitem
\bibitem[Power \latin{et~al.}(2011)Power, de~Menezes, Fagan, and
  Ferreira]{power2011magnetization}
Power,~S.~R.; de~Menezes,~V.; Fagan,~S.; Ferreira,~M. Magnetization profile for
  impurities in graphene nanoribbons. \emph{Phys. Rev. B} \textbf{2011},
  \emph{84}, 195431\relax
\mciteBstWouldAddEndPuncttrue
\mciteSetBstMidEndSepPunct{\mcitedefaultmidpunct}
{\mcitedefaultendpunct}{\mcitedefaultseppunct}\relax
\EndOfBibitem
\bibitem[Pi \latin{et~al.}(2009)Pi, McCreary, Bao, Han, Chiang, Li, Tsai, Lau,
  and Kawakami]{pi2009electronic}
Pi,~K.; McCreary,~K.; Bao,~W.; Han,~W.; Chiang,~Y.; Li,~Y.; Tsai,~S.-W.;
  Lau,~C.; Kawakami,~R. Electronic doping and scattering by transition metals
  on graphene. \emph{Phys. Rev. B} \textbf{2009}, \emph{80}, 075406\relax
\mciteBstWouldAddEndPuncttrue
\mciteSetBstMidEndSepPunct{\mcitedefaultmidpunct}
{\mcitedefaultendpunct}{\mcitedefaultseppunct}\relax
\EndOfBibitem
\bibitem[Krasheninnikov \latin{et~al.}(2009)Krasheninnikov, Lehtinen, Foster,
  Pyykk{\"o}, and Nieminen]{krasheninnikov2009embedding}
Krasheninnikov,~A.; Lehtinen,~P.; Foster,~A.~S.; Pyykk{\"o},~P.;
  Nieminen,~R.~M. Embedding transition-metal atoms in graphene: structure,
  bonding, and magnetism. \emph{Phys. Rev. Lett.} \textbf{2009}, \emph{102},
  126807\relax
\mciteBstWouldAddEndPuncttrue
\mciteSetBstMidEndSepPunct{\mcitedefaultmidpunct}
{\mcitedefaultendpunct}{\mcitedefaultseppunct}\relax
\EndOfBibitem
\bibitem[Santos \latin{et~al.}(2010)Santos, S{\'a}nchez-Portal, and
  Ayuela]{santos2010magnetism}
Santos,~E.~J.; S{\'a}nchez-Portal,~D.; Ayuela,~A. Magnetism of substitutional
  Co impurities in graphene: realization of single $\pi$ vacancies. \emph{Phys.
  Rev. B} \textbf{2010}, \emph{81}, 125433\relax
\mciteBstWouldAddEndPuncttrue
\mciteSetBstMidEndSepPunct{\mcitedefaultmidpunct}
{\mcitedefaultendpunct}{\mcitedefaultseppunct}\relax
\EndOfBibitem
\bibitem[Lisenkov \latin{et~al.}(2012)Lisenkov, Andriotis, and
  Menon]{lisenkov2012magnetic}
Lisenkov,~S.; Andriotis,~A.~N.; Menon,~M. Magnetic anisotropy and engineering
  of magnetic behavior of the edges in Co embedded graphene nanoribbons.
  \emph{Phys. Rev. Lett.} \textbf{2012}, \emph{108}, 187208\relax
\mciteBstWouldAddEndPuncttrue
\mciteSetBstMidEndSepPunct{\mcitedefaultmidpunct}
{\mcitedefaultendpunct}{\mcitedefaultseppunct}\relax
\EndOfBibitem
\bibitem[Boukhvalov and Katsnelson(2009)Boukhvalov, and
  Katsnelson]{boukhvalov2009destruction}
Boukhvalov,~D.; Katsnelson,~M. Destruction of graphene by metal adatoms.
  \emph{Appl. Phys. Lett.} \textbf{2009}, \emph{95}, 023109\relax
\mciteBstWouldAddEndPuncttrue
\mciteSetBstMidEndSepPunct{\mcitedefaultmidpunct}
{\mcitedefaultendpunct}{\mcitedefaultseppunct}\relax
\EndOfBibitem
\bibitem[Gyamfi \latin{et~al.}(2012)Gyamfi, Eelbo, Wa{\'s}niowska, Wehling,
  Forti, Starke, Lichtenstein, Katsnelson, and Wiesendanger]{gyamfi2012orbital}
Gyamfi,~M.; Eelbo,~T.; Wa{\'s}niowska,~M.; Wehling,~T.; Forti,~S.; Starke,~U.;
  Lichtenstein,~A.; Katsnelson,~M.; Wiesendanger,~R. Orbital selective coupling
  between Ni adatoms and graphene Dirac electrons. \emph{Phys. Rev. B}
  \textbf{2012}, \emph{85}, 161406\relax
\mciteBstWouldAddEndPuncttrue
\mciteSetBstMidEndSepPunct{\mcitedefaultmidpunct}
{\mcitedefaultendpunct}{\mcitedefaultseppunct}\relax
\EndOfBibitem
\bibitem[Eelbo \latin{et~al.}(2013)Eelbo, Wa{\'s}niowska, Gyamfi, Forti,
  Starke, and Wiesendanger]{eelbo2013influence}
Eelbo,~T.; Wa{\'s}niowska,~M.; Gyamfi,~M.; Forti,~S.; Starke,~U.;
  Wiesendanger,~R. Influence of the degree of decoupling of graphene on the
  properties of transition metal adatoms. \emph{Phys. Rev. B} \textbf{2013},
  \emph{87}, 205443\relax
\mciteBstWouldAddEndPuncttrue
\mciteSetBstMidEndSepPunct{\mcitedefaultmidpunct}
{\mcitedefaultendpunct}{\mcitedefaultseppunct}\relax
\EndOfBibitem
\bibitem[Donati \latin{et~al.}(2013)Donati, Dubout, Aut{\`e}s, Patthey,
  Calleja, Gambardella, Yazyev, and Brune]{donati2013magnetic}
Donati,~F.; Dubout,~Q.; Aut{\`e}s,~G.; Patthey,~F.; Calleja,~F.;
  Gambardella,~P.; Yazyev,~O.; Brune,~H. Magnetic moment and anisotropy of
  individual Co atoms on graphene. \emph{Phys. Rev. Lett.} \textbf{2013},
  \emph{111}, 236801\relax
\mciteBstWouldAddEndPuncttrue
\mciteSetBstMidEndSepPunct{\mcitedefaultmidpunct}
{\mcitedefaultendpunct}{\mcitedefaultseppunct}\relax
\EndOfBibitem
\bibitem[Valencia \latin{et~al.}(2010)Valencia, Gil, and
  Frapper]{valencia2010trends}
Valencia,~H.; Gil,~A.; Frapper,~G. Trends in the adsorption of 3d transition
  metal atoms onto graphene and nanotube surfaces: a DFT study and molecular
  orbital analysis. \emph{J. Phys. Chem. C} \textbf{2010}, \emph{114},
  14141--14153\relax
\mciteBstWouldAddEndPuncttrue
\mciteSetBstMidEndSepPunct{\mcitedefaultmidpunct}
{\mcitedefaultendpunct}{\mcitedefaultseppunct}\relax
\EndOfBibitem
\bibitem[Nguyen \latin{et~al.}(2017)Nguyen, Lin, Lin, Chiu, Tran, and
  Fa-Lin]{nguyen2017fluorination}
Nguyen,~D.~K.; Lin,~Y.-T.; Lin,~S.-Y.; Chiu,~Y.-H.; Tran,~N. T.~T.; Fa-Lin,~M.
  Fluorination-enriched electronic and magnetic properties in graphene
  nanoribbons. \emph{Phys. Chem. Chem. Phys.} \textbf{2017}, \emph{19},
  20667--20676\relax
\mciteBstWouldAddEndPuncttrue
\mciteSetBstMidEndSepPunct{\mcitedefaultmidpunct}
{\mcitedefaultendpunct}{\mcitedefaultseppunct}\relax
\EndOfBibitem
\bibitem[Gross and Dreizler(2013)Gross, and Dreizler]{gross2013density}
Gross,~E.~K.; Dreizler,~R.~M. \emph{Density functional theory}; Springer
  Science \& Business Media, 2013; Vol. 337\relax
\mciteBstWouldAddEndPuncttrue
\mciteSetBstMidEndSepPunct{\mcitedefaultmidpunct}
{\mcitedefaultendpunct}{\mcitedefaultseppunct}\relax
\EndOfBibitem
\bibitem[Lee \latin{et~al.}(2008)Lee, Lee, Ahn, Kim, Wilson, and
  John]{lee2008growth}
Lee,~J.-K.; Lee,~S.-C.; Ahn,~J.-P.; Kim,~S.-C.; Wilson,~J.~I.; John,~P. The
  growth of AA graphite on (111) diamond. \emph{J. Chem. Phys.} \textbf{2008},
  \emph{129}, 234709\relax
\mciteBstWouldAddEndPuncttrue
\mciteSetBstMidEndSepPunct{\mcitedefaultmidpunct}
{\mcitedefaultendpunct}{\mcitedefaultseppunct}\relax
\EndOfBibitem
\bibitem[Vilkov \latin{et~al.}(2013)Vilkov, Fedorov, Usachov, Yashina,
  Generalov, Borygina, Verbitskiy, Gruneis, and Vyalikh]{vilkov2013controlled}
Vilkov,~O.; Fedorov,~A.; Usachov,~D.; Yashina,~L.; Generalov,~A.; Borygina,~K.;
  Verbitskiy,~N.; Gruneis,~A.; Vyalikh,~D. Controlled assembly of
  graphene-capped nickel, cobalt and iron silicides. \emph{Sci. Rep.}
  \textbf{2013}, \emph{3}, 2168\relax
\mciteBstWouldAddEndPuncttrue
\mciteSetBstMidEndSepPunct{\mcitedefaultmidpunct}
{\mcitedefaultendpunct}{\mcitedefaultseppunct}\relax
\EndOfBibitem
\bibitem[Gao(2014)]{gao2014probing}
Gao,~L. Probing Electronic Properties of Graphene on the Atomic Scale by
  Scanning Tunneling Microscopy and Spectroscopy. \emph{Graphene and 2D
  Materials} \textbf{2014}, \emph{1}\relax
\mciteBstWouldAddEndPuncttrue
\mciteSetBstMidEndSepPunct{\mcitedefaultmidpunct}
{\mcitedefaultendpunct}{\mcitedefaultseppunct}\relax
\EndOfBibitem
\bibitem[Kresse and Furthm{\"u}ller(1996)Kresse, and
  Furthm{\"u}ller]{kresse1996efficient}
Kresse,~G.; Furthm{\"u}ller,~J. Efficient iterative schemes for ab initio
  total-energy calculations using a plane-wave basis set. \emph{Phys. Rev. B}
  \textbf{1996}, \emph{54}, 11169\relax
\mciteBstWouldAddEndPuncttrue
\mciteSetBstMidEndSepPunct{\mcitedefaultmidpunct}
{\mcitedefaultendpunct}{\mcitedefaultseppunct}\relax
\EndOfBibitem
\bibitem[Kresse and Joubert(1999)Kresse, and Joubert]{kresse1999ultrasoft}
Kresse,~G.; Joubert,~D. From ultrasoft pseudopotentials to the projector
  augmented-wave method. \emph{Phys. Rev. B} \textbf{1999}, \emph{59},
  1758\relax
\mciteBstWouldAddEndPuncttrue
\mciteSetBstMidEndSepPunct{\mcitedefaultmidpunct}
{\mcitedefaultendpunct}{\mcitedefaultseppunct}\relax
\EndOfBibitem
\bibitem[Bl{\"o}chl(1994)]{blochl1994projector}
Bl{\"o}chl,~P.~E. Projector augmented-wave method. \emph{Phys. Rev. B}
  \textbf{1994}, \emph{50}, 17953\relax
\mciteBstWouldAddEndPuncttrue
\mciteSetBstMidEndSepPunct{\mcitedefaultmidpunct}
{\mcitedefaultendpunct}{\mcitedefaultseppunct}\relax
\EndOfBibitem
\bibitem[Perdew \latin{et~al.}(1996)Perdew, Burke, and
  Ernzerhof]{perdew1996generalized}
Perdew,~J.~P.; Burke,~K.; Ernzerhof,~M. Generalized gradient approximation made
  simple. \emph{Phys. Rev. Lett.} \textbf{1996}, \emph{77}, 3865\relax
\mciteBstWouldAddEndPuncttrue
\mciteSetBstMidEndSepPunct{\mcitedefaultmidpunct}
{\mcitedefaultendpunct}{\mcitedefaultseppunct}\relax
\EndOfBibitem
\bibitem[Grimme(2006)]{grimme2006semiempirical}
Grimme,~S. Semiempirical GGA-type density functional constructed with a
  long-range dispersion correction. \emph{J. Theor. Comput. Chem.}
  \textbf{2006}, \emph{27}, 1787--1799\relax
\mciteBstWouldAddEndPuncttrue
\mciteSetBstMidEndSepPunct{\mcitedefaultmidpunct}
{\mcitedefaultendpunct}{\mcitedefaultseppunct}\relax
\EndOfBibitem
\bibitem[Mao \latin{et~al.}(2008)Mao, Yuan, and Zhong]{mao2008density}
Mao,~Y.; Yuan,~J.; Zhong,~J. Density functional calculation of transition metal
  adatom adsorption on graphene. \emph{J. Phys. Condens. Matter} \textbf{2008},
  \emph{20}, 115209\relax
\mciteBstWouldAddEndPuncttrue
\mciteSetBstMidEndSepPunct{\mcitedefaultmidpunct}
{\mcitedefaultendpunct}{\mcitedefaultseppunct}\relax
\EndOfBibitem
\bibitem[Tang \latin{et~al.}(2015)Tang, Chen, Li, Pan, Dai, and
  Ma]{tang2015adsorption}
Tang,~Y.; Chen,~W.; Li,~C.; Pan,~L.; Dai,~X.; Ma,~D. Adsorption behavior of Co
  anchored on graphene sheets toward NO, SO2, NH3, CO and HCN molecules.
  \emph{Appl. Surf. Sci.} \textbf{2015}, \emph{342}, 191--199\relax
\mciteBstWouldAddEndPuncttrue
\mciteSetBstMidEndSepPunct{\mcitedefaultmidpunct}
{\mcitedefaultendpunct}{\mcitedefaultseppunct}\relax
\EndOfBibitem
\bibitem[Eelbo \latin{et~al.}(2013)Eelbo, Wa{\'s}niowska, Thakur, Gyamfi,
  Sachs, Wehling, Forti, Starke, Tieg, Lichtenstein, \latin{et~al.}
  others]{eelbo2013adatoms}
Eelbo,~T.; Wa{\'s}niowska,~M.; Thakur,~P.; Gyamfi,~M.; Sachs,~B.; Wehling,~T.;
  Forti,~S.; Starke,~U.; Tieg,~C.; Lichtenstein,~A., \latin{et~al.}  Adatoms
  and clusters of 3d transition metals on graphene: Electronic and magnetic
  configurations. \emph{Phys. Rev. Lett.} \textbf{2013}, \emph{110},
  136804\relax
\mciteBstWouldAddEndPuncttrue
\mciteSetBstMidEndSepPunct{\mcitedefaultmidpunct}
{\mcitedefaultendpunct}{\mcitedefaultseppunct}\relax
\EndOfBibitem
\bibitem[Liu \latin{et~al.}(2014)Liu, Wang, Lin, Hupalo, Thiel, Ho, and
  Tringides]{liu2014structures}
Liu,~X.; Wang,~C.-Z.; Lin,~H.-Q.; Hupalo,~M.; Thiel,~P.~A.; Ho,~K.-M.;
  Tringides,~M.~C. Structures and magnetic properties of Fe clusters on
  graphene. \emph{Phys. Rev. B} \textbf{2014}, \emph{90}, 155444\relax
\mciteBstWouldAddEndPuncttrue
\mciteSetBstMidEndSepPunct{\mcitedefaultmidpunct}
{\mcitedefaultendpunct}{\mcitedefaultseppunct}\relax
\EndOfBibitem
\bibitem[Horing \latin{et~al.}(2013)Horing, Fessatidis, and
  Mancini]{horing2013atom}
Horing,~N. J.~M.; Fessatidis,~V.; Mancini,~J.~D. \emph{Low Dimensional
  Semiconductor Structures}; Springer, 2013; pp 93--99\relax
\mciteBstWouldAddEndPuncttrue
\mciteSetBstMidEndSepPunct{\mcitedefaultmidpunct}
{\mcitedefaultendpunct}{\mcitedefaultseppunct}\relax
\EndOfBibitem
\bibitem[Huang \latin{et~al.}(2014)Huang, Chen, Ho, Lin, and
  Lin]{huang2014feature}
Huang,~Y.-K.; Chen,~S.-C.; Ho,~Y.-H.; Lin,~C.-Y.; Lin,~M.-F. Feature-rich
  magnetic quantization in sliding bilayer graphenes. \emph{Sci. Rep.}
  \textbf{2014}, \emph{4}, 7509\relax
\mciteBstWouldAddEndPuncttrue
\mciteSetBstMidEndSepPunct{\mcitedefaultmidpunct}
{\mcitedefaultendpunct}{\mcitedefaultseppunct}\relax
\EndOfBibitem
\bibitem[Qiao \latin{et~al.}(2014)Qiao, Ren, Chen, Bellaiche, Zhang, MacDonald,
  and Niu]{qiao2014quantum}
Qiao,~Z.; Ren,~W.; Chen,~H.; Bellaiche,~L.; Zhang,~Z.; MacDonald,~A.; Niu,~Q.
  Quantum anomalous Hall effect in graphene proximity coupled to an
  antiferromagnetic insulator. \emph{Phys. Rev. Lett.} \textbf{2014},
  \emph{112}, 116404\relax
\mciteBstWouldAddEndPuncttrue
\mciteSetBstMidEndSepPunct{\mcitedefaultmidpunct}
{\mcitedefaultendpunct}{\mcitedefaultseppunct}\relax
\EndOfBibitem
\bibitem[Santoso \latin{et~al.}(2014)Santoso, Singh, Gogoi, Asmara, Wei, Chen,
  Wee, Pereira, and Rusydi]{santoso2014tunable}
Santoso,~I.; Singh,~R.~S.; Gogoi,~P.~K.; Asmara,~T.~C.; Wei,~D.; Chen,~W.;
  Wee,~A.~T.; Pereira,~V.~M.; Rusydi,~A. Tunable optical absorption and
  interactions in graphene via oxygen plasma. \emph{Phys. Rev. B}
  \textbf{2014}, \emph{89}, 075134\relax
\mciteBstWouldAddEndPuncttrue
\mciteSetBstMidEndSepPunct{\mcitedefaultmidpunct}
{\mcitedefaultendpunct}{\mcitedefaultseppunct}\relax
\EndOfBibitem
\bibitem[Chen \latin{et~al.}(2014)Chen, Chiu, Wu, and Lin]{chen2014shift}
Chen,~S.-C.; Chiu,~C.-W.; Wu,~C.-L.; Lin,~M.-F. Shift-enriched optical
  properties in bilayer graphene. \emph{RSC Adv.} \textbf{2014}, \emph{4},
  63779--63783\relax
\mciteBstWouldAddEndPuncttrue
\mciteSetBstMidEndSepPunct{\mcitedefaultmidpunct}
{\mcitedefaultendpunct}{\mcitedefaultseppunct}\relax
\EndOfBibitem
\bibitem[Johll \latin{et~al.}(2014)Johll, Lee, Ng, Kang, and
  Tok]{johll2014influence}
Johll,~H.; Lee,~M. D.~K.; Ng,~S. P.~N.; Kang,~H.~C.; Tok,~E.~S. Influence of
  interconfigurational electronic states on Fe, Co, Ni-silicene materials
  selection for spintronics. \emph{Sci. Rep.} \textbf{2014}, \emph{4},
  7594\relax
\mciteBstWouldAddEndPuncttrue
\mciteSetBstMidEndSepPunct{\mcitedefaultmidpunct}
{\mcitedefaultendpunct}{\mcitedefaultseppunct}\relax
\EndOfBibitem
\bibitem[Sprinkle \latin{et~al.}(2009)Sprinkle, \latin{et~al.}
  others]{sprinkle2009first}
Sprinkle,~M., \latin{et~al.}  First direct observation of a nearly ideal
  graphene band structure. \emph{Phys. Rev. Lett.} \textbf{2009}, \emph{103},
  226803\relax
\mciteBstWouldAddEndPuncttrue
\mciteSetBstMidEndSepPunct{\mcitedefaultmidpunct}
{\mcitedefaultendpunct}{\mcitedefaultseppunct}\relax
\EndOfBibitem
\bibitem[Virojanadara \latin{et~al.}(2010)Virojanadara, Watcharinyanon,
  Zakharov, and Johansson]{virojanadara2010epitaxial}
Virojanadara,~C.; Watcharinyanon,~S.; Zakharov,~A.; Johansson,~L.~I. Epitaxial
  graphene on 6 H-SiC and Li intercalation. \emph{Phys. Rev. B} \textbf{2010},
  \emph{82}, 205402\relax
\mciteBstWouldAddEndPuncttrue
\mciteSetBstMidEndSepPunct{\mcitedefaultmidpunct}
{\mcitedefaultendpunct}{\mcitedefaultseppunct}\relax
\EndOfBibitem
\bibitem[Sugawara \latin{et~al.}(2011)Sugawara, Kanetani, Sato, and
  Takahashi]{sugawara2011fabrication}
Sugawara,~K.; Kanetani,~K.; Sato,~T.; Takahashi,~T. Fabrication of
  Li-intercalated bilayer graphene. \emph{AIP Advances} \textbf{2011},
  \emph{1}, 022103\relax
\mciteBstWouldAddEndPuncttrue
\mciteSetBstMidEndSepPunct{\mcitedefaultmidpunct}
{\mcitedefaultendpunct}{\mcitedefaultseppunct}\relax
\EndOfBibitem
\bibitem[Bostwick \latin{et~al.}(2007)Bostwick, Ohta, Seyller, Horn, and
  Rotenberg]{bostwick2007quasiparticle}
Bostwick,~A.; Ohta,~T.; Seyller,~T.; Horn,~K.; Rotenberg,~E. Quasiparticle
  dynamics in graphene. \emph{Nat. Phys.} \textbf{2007}, \emph{3}, 36\relax
\mciteBstWouldAddEndPuncttrue
\mciteSetBstMidEndSepPunct{\mcitedefaultmidpunct}
{\mcitedefaultendpunct}{\mcitedefaultseppunct}\relax
\EndOfBibitem
\bibitem[Ohta \latin{et~al.}(2006)Ohta, Bostwick, Seyller, Horn, and
  Rotenberg]{ohta2006controlling}
Ohta,~T.; Bostwick,~A.; Seyller,~T.; Horn,~K.; Rotenberg,~E. Controlling the
  electronic structure of bilayer graphene. \emph{Science} \textbf{2006},
  \emph{313}, 951--954\relax
\mciteBstWouldAddEndPuncttrue
\mciteSetBstMidEndSepPunct{\mcitedefaultmidpunct}
{\mcitedefaultendpunct}{\mcitedefaultseppunct}\relax
\EndOfBibitem
\bibitem[Bornemann \latin{et~al.}(2012)Bornemann, {\v{S}}ipr, Mankovsky,
  Polesya, Staunton, Wurth, Ebert, and Min{\'a}r]{bornemann2012trends}
Bornemann,~S.; {\v{S}}ipr,~O.; Mankovsky,~S.; Polesya,~S.; Staunton,~J.;
  Wurth,~W.; Ebert,~H.; Min{\'a}r,~J. Trends in the magnetic properties of Fe,
  Co, and Ni clusters and monolayers on Ir (111), Pt (111), and Au (111).
  \emph{Phys. Rev. B} \textbf{2012}, \emph{86}, 104436\relax
\mciteBstWouldAddEndPuncttrue
\mciteSetBstMidEndSepPunct{\mcitedefaultmidpunct}
{\mcitedefaultendpunct}{\mcitedefaultseppunct}\relax
\EndOfBibitem
\bibitem[Paudyal \latin{et~al.}(2004)Paudyal, Saha-Dasgupta, and
  Mookerjee]{paudyal2004magnetic}
Paudyal,~D.; Saha-Dasgupta,~T.; Mookerjee,~A. Magnetic properties of X--Pt (X=
  Fe, Co, Ni) alloy systems. \emph{J. Phys-Condens. Mat.} \textbf{2004},
  \emph{16}, 2317\relax
\mciteBstWouldAddEndPuncttrue
\mciteSetBstMidEndSepPunct{\mcitedefaultmidpunct}
{\mcitedefaultendpunct}{\mcitedefaultseppunct}\relax
\EndOfBibitem
\bibitem[Zhang \latin{et~al.}(2010)Zhang, Chen, Wang, and
  Xu]{zhang2010comparison}
Zhang,~J.-M.; Chen,~L.-Y.; Wang,~S.-F.; Xu,~K.-W. Comparison of the structural,
  electronic and magnetic properties of Fe, Co and Ni nanowires encapsulated
  into silicon carbide nanotube. \emph{Eur. Phys. J. B} \textbf{2010},
  \emph{73}, 555--561\relax
\mciteBstWouldAddEndPuncttrue
\mciteSetBstMidEndSepPunct{\mcitedefaultmidpunct}
{\mcitedefaultendpunct}{\mcitedefaultseppunct}\relax
\EndOfBibitem
\bibitem[Serrate \latin{et~al.}(2010)Serrate, Ferriani, Yoshida, Hla, Menzel,
  Von~Bergmann, Heinze, Kubetzka, and Wiesendanger]{serrate2010imaging}
Serrate,~D.; Ferriani,~P.; Yoshida,~Y.; Hla,~S.-W.; Menzel,~M.;
  Von~Bergmann,~K.; Heinze,~S.; Kubetzka,~A.; Wiesendanger,~R. Imaging and
  manipulating the spin direction of individual atoms. \emph{Nat. Nanotechnol.}
  \textbf{2010}, \emph{5}, 350--353\relax
\mciteBstWouldAddEndPuncttrue
\mciteSetBstMidEndSepPunct{\mcitedefaultmidpunct}
{\mcitedefaultendpunct}{\mcitedefaultseppunct}\relax
\EndOfBibitem
\bibitem[Wulfhekel and Kirschner(2007)Wulfhekel, and
  Kirschner]{wulfhekel2007spin}
Wulfhekel,~W.; Kirschner,~J. Spin-polarized scanning tunneling microscopy of
  magnetic structures and antiferromagnetic thin films. \emph{Mater. Res.}
  \textbf{2007}, \emph{37}, 69\relax
\mciteBstWouldAddEndPuncttrue
\mciteSetBstMidEndSepPunct{\mcitedefaultmidpunct}
{\mcitedefaultendpunct}{\mcitedefaultseppunct}\relax
\EndOfBibitem
\bibitem[Neto \latin{et~al.}(2009)Neto, Guinea, Peres, Novoselov, and
  Geim]{neto2009electronic}
Neto,~A.~C.; Guinea,~F.; Peres,~N.~M.; Novoselov,~K.~S.; Geim,~A.~K. The
  electronic properties of graphene. \emph{Rev. Mod. Phys.} \textbf{2009},
  \emph{81}, 109\relax
\mciteBstWouldAddEndPuncttrue
\mciteSetBstMidEndSepPunct{\mcitedefaultmidpunct}
{\mcitedefaultendpunct}{\mcitedefaultseppunct}\relax
\EndOfBibitem
\bibitem[Li \latin{et~al.}(2009)Li, Luican, and Andrei]{li2009scanning}
Li,~G.; Luican,~A.; Andrei,~E.~Y. Scanning tunneling spectroscopy of graphene
  on graphite. \emph{Phys. Rev. Lett.} \textbf{2009}, \emph{102}, 176804\relax
\mciteBstWouldAddEndPuncttrue
\mciteSetBstMidEndSepPunct{\mcitedefaultmidpunct}
{\mcitedefaultendpunct}{\mcitedefaultseppunct}\relax
\EndOfBibitem
\bibitem[Chen \latin{et~al.}(2015)Chen, Su, Chang, Cheng, Chong, Huang, and
  Lin]{chen2015long}
Chen,~H.-H.; Su,~S.; Chang,~S.-L.; Cheng,~B.-Y.; Chong,~C.-W.; Huang,~J.;
  Lin,~M.-F. Long-range interactions of bismuth growth on monolayer epitaxial
  graphene at room temperature. \emph{Carbon} \textbf{2015}, \emph{93},
  180--186\relax
\mciteBstWouldAddEndPuncttrue
\mciteSetBstMidEndSepPunct{\mcitedefaultmidpunct}
{\mcitedefaultendpunct}{\mcitedefaultseppunct}\relax
\EndOfBibitem
\bibitem[Cort{\'e}s-Arriagada \latin{et~al.}(2018)Cort{\'e}s-Arriagada,
  Villegas-Escobar, and Ortega]{cortes2018fe}
Cort{\'e}s-Arriagada,~D.; Villegas-Escobar,~N.; Ortega,~D.~E. Fe-doped graphene
  nanosheet as an adsorption platform of harmful gas molecules (CO, CO2, SO2
  and H2S), and the co-adsorption in O2 environments. \emph{Appl. Surf. Sci.}
  \textbf{2018}, \emph{427}, 227--236\relax
\mciteBstWouldAddEndPuncttrue
\mciteSetBstMidEndSepPunct{\mcitedefaultmidpunct}
{\mcitedefaultendpunct}{\mcitedefaultseppunct}\relax
\EndOfBibitem
\bibitem[Cort{\'e}s-Arriagada \latin{et~al.}(2017)Cort{\'e}s-Arriagada,
  Villegas-Escobar, Miranda-Rojas, and Toro-Labb{\'e}]{cortes2017adsorption}
Cort{\'e}s-Arriagada,~D.; Villegas-Escobar,~N.; Miranda-Rojas,~S.;
  Toro-Labb{\'e},~A. Adsorption/desorption process of formaldehyde onto iron
  doped graphene: a theoretical exploration from density functional theory
  calculations. \emph{Phys. Chem. Chem. Phys.} \textbf{2017}, \emph{19},
  4179--4189\relax
\mciteBstWouldAddEndPuncttrue
\mciteSetBstMidEndSepPunct{\mcitedefaultmidpunct}
{\mcitedefaultendpunct}{\mcitedefaultseppunct}\relax
\EndOfBibitem
\bibitem[Wanno and Tabtimsai(2014)Wanno, and Tabtimsai]{wanno2014dft}
Wanno,~B.; Tabtimsai,~C. A DFT investigation of CO adsorption on VIIIB
  transition metal-doped graphene sheets. \emph{Superlattices and
  Microstructures} \textbf{2014}, \emph{67}, 110--117\relax
\mciteBstWouldAddEndPuncttrue
\mciteSetBstMidEndSepPunct{\mcitedefaultmidpunct}
{\mcitedefaultendpunct}{\mcitedefaultseppunct}\relax
\EndOfBibitem
\bibitem[Zhang \latin{et~al.}(2014)Zhang, Luo, Song, Lin, Lu, and
  Tang]{zhang2014dft}
Zhang,~H.-p.; Luo,~X.-g.; Song,~H.-t.; Lin,~X.-y.; Lu,~X.; Tang,~Y. DFT study
  of adsorption and dissociation behavior of H2S on Fe-doped graphene.
  \emph{Appl. Surf. Sci.} \textbf{2014}, \emph{317}, 511--516\relax
\mciteBstWouldAddEndPuncttrue
\mciteSetBstMidEndSepPunct{\mcitedefaultmidpunct}
{\mcitedefaultendpunct}{\mcitedefaultseppunct}\relax
\EndOfBibitem
\bibitem[Cort{\'e}s-Arriagada(2016)]{cortes2016expanding}
Cort{\'e}s-Arriagada,~D. Expanding the environmental applications of metal (Al,
  Ti, Mn, Fe) doped graphene: adsorption and removal of 1, 4-dioxane.
  \emph{Phys. Chem. Chem. Phys.} \textbf{2016}, \emph{18}, 32281--32292\relax
\mciteBstWouldAddEndPuncttrue
\mciteSetBstMidEndSepPunct{\mcitedefaultmidpunct}
{\mcitedefaultendpunct}{\mcitedefaultseppunct}\relax
\EndOfBibitem
\bibitem[Chen \latin{et~al.}(2015)Chen, Yang, Zhou, Lai, Rauf, Wang, Pan,
  Zhuang, Wang, Wang, \latin{et~al.} others]{chen2015aminothiazole}
Chen,~C.; Yang,~X.-D.; Zhou,~Z.-Y.; Lai,~Y.-J.; Rauf,~M.; Wang,~Y.; Pan,~J.;
  Zhuang,~L.; Wang,~Q.; Wang,~Y.-C., \latin{et~al.}  Aminothiazole-derived N,
  S, Fe-doped graphene nanosheets as high performance electrocatalysts for
  oxygen reduction. \emph{Chemical Communications} \textbf{2015}, \emph{51},
  17092--17095\relax
\mciteBstWouldAddEndPuncttrue
\mciteSetBstMidEndSepPunct{\mcitedefaultmidpunct}
{\mcitedefaultendpunct}{\mcitedefaultseppunct}\relax
\EndOfBibitem
\bibitem[Wu \latin{et~al.}(2010)Wu, Ren, Wen, Gao, Zhao, Chen, Zhou, Li, and
  Cheng]{wu2010graphene}
Wu,~Z.-S.; Ren,~W.; Wen,~L.; Gao,~L.; Zhao,~J.; Chen,~Z.; Zhou,~G.; Li,~F.;
  Cheng,~H.-M. Graphene anchored with Co3O4 nanoparticles as anode of lithium
  ion batteries with enhanced reversible capacity and cyclic performance.
  \emph{ACS nano} \textbf{2010}, \emph{4}, 3187--3194\relax
\mciteBstWouldAddEndPuncttrue
\mciteSetBstMidEndSepPunct{\mcitedefaultmidpunct}
{\mcitedefaultendpunct}{\mcitedefaultseppunct}\relax
\EndOfBibitem
\bibitem[Hassoun \latin{et~al.}(2014)Hassoun, Bonaccorso, Agostini, Angelucci,
  Betti, Cingolani, Gemmi, Mariani, Panero, Pellegrini, \latin{et~al.}
  others]{hassoun2014advanced}
Hassoun,~J.; Bonaccorso,~F.; Agostini,~M.; Angelucci,~M.; Betti,~M.~G.;
  Cingolani,~R.; Gemmi,~M.; Mariani,~C.; Panero,~S.; Pellegrini,~V.,
  \latin{et~al.}  An advanced lithium-ion battery based on a graphene anode and
  a lithium iron phosphate cathode. \emph{Nano Lett.} \textbf{2014}, \emph{14},
  4901--4906\relax
\mciteBstWouldAddEndPuncttrue
\mciteSetBstMidEndSepPunct{\mcitedefaultmidpunct}
{\mcitedefaultendpunct}{\mcitedefaultseppunct}\relax
\EndOfBibitem
\bibitem[Atabaki and Kovacevic(2013)Atabaki, and
  Kovacevic]{atabaki2013graphene}
Atabaki,~M.~M.; Kovacevic,~R. Graphene composites as anode materials in
  lithium-ion batteries. \emph{Electronic Materials Letters} \textbf{2013},
  \emph{9}, 133--153\relax
\mciteBstWouldAddEndPuncttrue
\mciteSetBstMidEndSepPunct{\mcitedefaultmidpunct}
{\mcitedefaultendpunct}{\mcitedefaultseppunct}\relax
\EndOfBibitem
\bibitem[Yu \latin{et~al.}(2016)Yu, Qu, Zhao, Li, Chen, Sun, Gao, and
  Zhu]{yu2016growth}
Yu,~X.; Qu,~B.; Zhao,~Y.; Li,~C.; Chen,~Y.; Sun,~C.; Gao,~P.; Zhu,~C. Growth of
  Hollow Transition Metal (Fe, Co, Ni) Oxide Nanoparticles on Graphene Sheets
  through Kirkendall Effect as Anodes for High-Performance Lithium-Ion
  Batteries. \emph{Chem. A Eur. J.} \textbf{2016}, \emph{22}, 1638--1645\relax
\mciteBstWouldAddEndPuncttrue
\mciteSetBstMidEndSepPunct{\mcitedefaultmidpunct}
{\mcitedefaultendpunct}{\mcitedefaultseppunct}\relax
\EndOfBibitem
\end{mcitethebibliography}
	
\end{document}